\documentclass[twocolumn,numberedappendix]{openjournal}

\usepackage{tikz}
\usepackage{overpic}
\usepackage{graphicx}
\usepackage{rotating}
\usepackage{multirow}
\usepackage{colorblind}
\usetikzlibrary{calc}
\usetikzlibrary{fit}
\usepackage{amsmath}
\usepackage{xcolor}
\usepackage{hyperref}
\usepackage{orcidlink}
\definecolor{xlinkcolor}{cmyk}{1,1,0,0}
\hypersetup{colorlinks=true,linkcolor=xlinkcolor,citecolor=xlinkcolor,filecolor=xlinkcolor,urlcolor=xlinkcolor}
\makeatletter
\def\frontmatter@title@below{\vspace*{-2.63\baselineskip\vspace*{0.05in}}}
\makeatother

\journalinfo{The Open Journal of Astrophysics}
\submitted{}

\begin{document}

\title{High-Precision Differential Radial Velocities of C3PO Wide Binaries: \\ A Test of Modified Newtonian Dynamics (MOND)}

\author{Serat M. Saad\,\orcidlink{0009-0004-9592-2311}}
\affiliation{Department of Astronomy, The Ohio State University, 140 W.\,18th Ave., Columbus, OH 43210, USA}
\email{saad.104@osu.edu}
\affiliation{Center for Cosmology and AstroParticle Physics, The Ohio State University, 140 W. 18th Ave, Columbus, OH, 43210, USA}

\author{Yuan-Sen Ting\,\orcidlink{0000-0001-5082-9536}}
\affiliation{Department of Astronomy, The Ohio State University, 140 W.\,18th Ave., Columbus, OH 43210, USA}
\affiliation{Center for Cosmology and AstroParticle Physics, The Ohio State University, 140 W. 18th Ave, Columbus, OH, 43210, USA}
\affiliation{Max-Planck-Institut f\"{u}r Astronomie, K\"{o}nigstuhl 17, D-69117 Heidelberg, Germany}
\email{ting.74@osu.edu}

\begin{abstract}
Wide-binary stars, separated by thousands of AU, reside in low-acceleration regimes where Modified Newtonian Dynamics (MOND) predicts deviation from Newtonian gravity. However, \textit{Gaia} radial velocities (RVs) lack the precision to resolve the small velocity differences expected in these systems, limiting previous MOND analyses to two-dimensional kinematics. In this paper, we introduce a technique to measure differential RVs of wide binary stars using high resolution, high signal-to-noise spectra. We apply this method to measure differential RVs of 85 wide-binaries from the C3PO survey and achieved precisions of $\sim$8--15 m\,s$^{-1}$ per binary pair, a $\sim$10--100$\times$ improvement (median $\sim24\times$) over \textit{Gaia} DR3. After applying a selection criterion based on the scaled velocity, we retain 57 gravitationally bound systems for further analysis. Combining these measurements with \textit{Gaia} astrometry, we construct a hierarchical Bayesian model to infer the orbital elements of all wide-binary pairs and the global MOND acceleration scale ($a_0$). We test two commonly used interpolating functions in MOND formulation: the simple form ($b=1$, $\mu = x/(1+x)$) and the standard form ($b=2$, $\mu = x/\sqrt{1+x^2}$). 
Our results indicate tension with MOND at the presently accepted $a_0$ value: for $b=1$, the canonical value is excluded at $2.5\sigma$, while for $b=2$, the exclusion is at $1.5\sigma$. We discuss how systematic uncertainties in the external field effect treatment, particularly in the transition regime, may affect these conclusions.
\end{abstract}


\section{Introduction}

The physics of gravity in low acceleration regime remains one of the unsolved problems in science. Since the last century, evidence has indicated that the dynamics of galaxies and large-scale structures cannot be explained by visible matter alone under Newtonian gravity \citep{Zwicky1937, Rubin1970, Rubin1980,Begeman1989, Sofue2001}. To address this discrepancy, the standard cosmological paradigm ($\lambda$CDM) includes Cold Dark Matter (CDM) as the component which constitutes most of the mass of galaxies and galaxy clusters. This framework has achieved a lot of successes like accurate predictions of the cosmic microwave background power spectrum, large-scale structure formation, and baryon acoustic oscillation \citep{Planck2020a, Planck2020b,Springel2005, Eisenstein2005}.   

Nonetheless, to date, no direct detection of dark matter particles has been achieved. Furthermore, several observations on galactic scales present difficulties for the $\Lambda$CDM framework, like the core-cusp problem, the missing satellite problem, and the tight correlations between baryonic and dynamical properties of galaxies \citep{Weinberg2015, Bullock2017}. Although, many of these anomalies can potentially be explained by baryon physics \citep{Pontzen2012,DiCinito2014,Wetzel2016,Hopkins2018}.

As an alternative to dark matter, \cite{Milgrom1983} proposed  Modified Newtonian Dynamics (MOND). MOND provides a modification of Newtonian gravity in the regime where accelerations fall below a scale of $a_0 \approx 1.2 \times 10^{-10} \rm \: ms^{-2}$ \citep{Milgrom1983, Milgrom1983b}. MOND can reproduce the flat rotation curves of spiral galaxies, the baryonic Tully-Fisher relation, and the internal dynamics of low-surface-brightness galaxies using only the universal parameter $a_0$ and without invoking dark matter \citep{Sanders2002, Famaey2012, Mcgaugh2016, Lelli2016}. Notably, the slope of the baryonic Tully-Fisher relation was predicted by MOND in the original papers by \cite{Milgrom1983, Milgrom1983b}, before it was observationally confirmed \citep{McGaugh2000}. To predict gravitational lensing and cosmological observables properly, MOND was later embedded in Lagrangian-based frameworks like AQUAL \citep{Bekenstein1984} and relativistic extensions like TeVes \citep{Bekenstein2004}.

However, MOND also faces challenges at the scale of galaxy clusters and cosmology where additional mass beyond baryons seems required even in MOND framework \citep{Sanders2003, Angus2008}. This has motivated tests of MOND in environments different from galactic rotation curves, like wide binary stars in the solar neighborhood \citep{Hernandez2012, Banik2018, Pittordis2018}. These wide binary stars, separated by thousands of AU, have gravitational acceleration in the scale of $a_0$, which is a suitable low acceleration regime to test MOND. On top of that, wide binaries should contain negligible dark matter even in the CDM paradigm, as the local dark matter density contributes only one part in $\sim 10^5$ of the binary's total mass within the orbit \citep{Hernandez2023}. This makes testing MOND with wide-binaries independent of dark matter assumptions.

Since its inception, \textit{Gaia} has provided precise astrometric data of billions of stars \citep{Gaia2016, Gaia2021, Gaia2023}. This has enabled the identification of large samples of co-moving wide-binary pairs through their sky-projected relative velocities \citep{Oh2017, Elbadry2018, Hartman2020, Tian2020, ElBadry2021}. Several initiatives have been taken to use \textit{Gaia} wide binaries data to test MOND, which ended up in conflicting results. \cite{Hernandez2022, Hernandez2023, Hernandez2024} and \cite{Chae2023, Chae2024a, Chae2024b} have found evidence for gravitational anomaly consistent with MOND predictions at separations beyond $\sim$2--3 kAU. On the other hand, \cite{Pittordis2023} and \cite{Banik2024} found the wide binary systems follow Newtonian gravity more than MOND. As discussed in the review by \cite{Hernandez2024review}, several factors contribute to these conflicting results, including differences in sample selection and treatment of hidden triple systems. 

A limitation of existing tests is their reliance on sky-projected relative velocities derived from \textit{Gaia} proper motions. Though \textit{Gaia} also provides radial velocities (RVs), their uncertainties are typically $\sim1000~\rm{m\,s^{-1}}$ \citep{Katz2023}, exceeding the expected velocity differences of $\sim50$--$200~\rm{m\,s^{-1}}$ in these wide binaries. Such large uncertainties lack the statistical power to distinguish between MOND and Newtonian model. Beyond this, the absence of precise RVs prevents constraints on the full three-dimensional kinematics of these systems. 

As discussed by \cite{El-badry2019}, comparing the distribution of observed velocities to forward-simulated predictions under different gravity models has limited ability to validate individual systems, identify outliers, or propagate uncertainties from orbital degeneracies. In such case, a hierarchical Bayesian inference is necessary to fit orbital elements for each system along with global parameters like $a_0$. But this requires precise RV measurement to constrain individual orbits.

The theoretical precision limit for RV measurements can be estimated from the Cram{\'e}r-Rao bound \citep{rao1945information}. For a spectrum with flux $F(\lambda)$, the minimum RV uncertainty scales as $\sigma_v \propto ({\rm SNR})^{-1} (\sum_i |\partial F_i/\partial \lambda|^2)^{-1/2}$, where the sum runs over spectral pixels \citep{Bouchy2001}. For \textit{Gaia} RVs spectra ($R \approx 11500$, SNR $\sim 20$--50 per transit, wavelength coverage of 845--872 nm), this yields uncertainties of $\sim$500--2000 m\,s$^{-1}$ for solar-type stars. In contrast, high-resolution echelle spectra ($R\approx50000$, SNR $\sim$150--300, spanning 3500--9000 \AA) contain more spectral information. Our observations improve on all three factors that determine RV precision: spectral resolution $R$, SNR, and wavelength coverage (the number of spectral lines), achieving final differential RV precisions of $\sim$8--15 m\,s$^{-1}$ per binary pair (see Section~\ref{sec: Data}).

The Complete Census of Co-moving Pairs of Stars (C3PO) survey was designed to collect the largest sample of high resolution, high signal-to-noise ratio (SNR) spectra of co-moving wide-binaries to date \citep{C3POI}. Previous C3PO projects searched for planet engulfment in co-natal wide-binaries, where one star served as a chemical control for the other \citep{C3POII, C3POIII, C3POIV, C3POV}. Particularly, C3PO II \citep{C3POII} applied Bayesian model comparison to test whether observed abundance patterns were more consistent with a flat baseline, atomic diffusion, or planetary ingestion. This analysis identified eight planet-engulfment candidates among all the C3PO sources. Though C3PO was originally motivated by chemical abundance science, the same high-resolution, high SNR spectra also enable precise RV measurements. These precise RVs allows for a precision test of MOND.

In this paper, we test MOND using differential RV measurements of C3PO wide binaries. In Section~\ref{sec: Data}, we describe the C3PO dataset and the differential RV measurement technique. In Section~\ref{sec:formulation}, we present the MOND formulation, geometry, and all the observable quantities that we adopt for this work. In Section~\ref{sec: Methods}, we present the model framework, inference steps, and the posterior results of our analysis. Finally, in Section~\ref{sec: Discussion}, we discuss the implications of our findings along with their relevance to the ongoing efforts to test MOND, and possible future works.

\section{Data \& Measurements} \label{sec: Data}
\subsection{Data}
The C3PO survey collected high resolution, high SNR spectra of 125 co-moving star pairs (250 stars in total). To construct the C3PO sample spatial separations and 3D velocity separations were first measured for co-moving pairs in \textit{Gaia} DR3 \citep{Gaia2023} data using a similar approach to \cite{Nelson2021}. Constraints were then applied to spatial separations ($\Delta s$ $< 10^{6.8} \; \rm AU$) and 3D velocity separations ($\Delta v < 2 \text{ km s}^{-1}$) to limit contamination from non-co-natal pairs. Then additional photometric cuts were applied in the following way: (i) Color ($0.65 \leq \rm BP - RP \leq 1.15 \: mag$ and $|\Delta \rm BP-RP| < 0.15 \: mag$), to ensure the homogeneity in the stellar type; (ii) Brightness ($\rm G \:mag < 10 \: mag$), to ensure high SNR with low telescope time for precise spectroscopic measurement; and (iii) Absolute Magnitude ($\Delta M_{G} < 1 \: \rm mag$), to ensure similarity in intrinsic luminosity between the binary pairs. Because of the color selection, the resulting C3PO sample is mostly consists of F, G, and K dwarfs with possible mass range from $0.8$--$1.2~\rm M_\odot$. The metallicity distribution spans approximately $-0.5$ to $+0.3$ dex, which represents the nearby thin-disk population. Because all C3PO targets are bright ($G < 10$ mag), the per-pixel SNR ranges from $150$ to $300$. We also used astrometry for all the C3PO sources from \textit{Gaia}, including parallaxes, proper motion, and RVs with their uncertainties. Though we did not use the \textit{Gaia} RVs directly, we used them to verify consistency with our measurements.

All C3PO sources were observed using three high-resolution spectrographs: the Magellan Inamori Kyocera Echelle (MIKE) spectrograph on the Magellan Telescope over 7 nights ($R \approx 50000$), Ultraviolet and Visual Echelle Spectrograph (UVES) on the European Southern Observatory's Very Large Telescope over $26.4 \text{ h}$ ($R \approx 50000$), and High Resolution Echelle Spectrometer (HIRES) on the Keck Telescope over 1 night ($R \approx 72000$). A total of 78 pairs were observed using Magellan, 25 pairs with VLT, and 22 pairs with Keck. MIKE spectra were reduced using the \texttt{CarPy} pipeline; UVES spectra were reduced with the \texttt{ESO Reflex} workflow; and HIRES data were obtained in pipeline-reduced form from the Keck Observatory Archive \citep{Bernstein2003MIKE, Dodorico2000UVES, Vogt1994HiRes}. For all instruments, we used wavelength calibrated, blaze-corrected, order-by-order extracted 1-D spectra produced by the respective pipelines. From the original 125 pairs, we excluded 22 Keck/HIRES pairs because pipeline-reduced 1-D echelle spectra were not available for those systems. An additional two Magellan/MIKE pairs and one VLT/UVES pair were removed because their extracted 1-D spectra were too noisy for reliable differential RV measurements and would require re-reduction from raw frames. Finally, we excluded fifteen additional pairs for which at least one component lacked a Gaia FLAME mass estimate \citep{Gaiaflame, Gaiaflame2}. This resulted in a working sample of 85 pairs for differential RV measurements. Gaia FLAME provides stellar masses by fitting multi-band photometry and low-resolution spectra to grids of stellar evolution models, that is isochrone fitting in color-magnitude and spectral space \citep{Gaiaflame2}. We use FLAME masses throughout this work as priors in our orbital model (Section~\ref{sec: Methods}), and also for the bound system selection described below. The typical FLAME mass uncertainties for our sample of bright F, G, and K dwarfs are $\sim$0.05--0.10~${\rm M}_\odot$.

\subsection{Differential Radial Velocity}

Measuring absolute RVs at the $\sim10~\rm m\,s^{-1}$ level is challenging due to systematics including wavelength calibration errors, spectrograph drift, and astrophysical effects such as convective blueshift and gravitational redshift. For wide binaries observed with the same instrument under similar conditions, many of these systematics cancel when measuring differential RVs between the two components. This has a similar philosophy to the differential abundance technique in stellar spectroscopy, where comparing chemically similar stars observed in identical conditions yields more precise relative abundances \citep{Melendez2009, Bedell2018}.

To measure differential RVs between wide-binary components, we developed a forward-modeling approach that accounts for the pixel-integrated nature of the echelle spectroscopy.  Rather than cross-correlating flux samples directly, we designated one star in each pair as the template and the other as the comparison. For each echelle order of the template spectrum, we fitted a cubic spline. From this spline we computed a cumulative flux function $C(\lambda)$ to rapidly evaluate pixel-integrated model fluxes. For a trial velocity shift $v$, the comparison wavelength are related to the template wavelengths through a Doppler factor, $s = 1 + v/c$. The model flux for an observed pixel with wavelength boundaries $[a_i,\, b_i]$ is then,

\begin{equation}
    F^{\rm mod}_i(v) = s\,\frac{C(b_i/s) - C(a_i/s)}{b_i - a_i}
\end{equation}
This provides the photon-weighted flux, integrated over each detector pixel. We then evaluated a chi-squared statistic:

\begin{equation}
    \chi^2(v) = \sum_i W_i\,\left[F^{\rm obs}_i - F^{\rm mod}_i(v)\right]^2
\end{equation}
where $W_i$ are inverse-variance weights. This chi-squared was calculated over a velocity grid from $-5$ to $+5~\mathrm{km\,s^{-1}}$ in steps of $40~\mathrm{m\,s^{-1}}$. This grid spacing of $40~\mathrm{m\,s^{-1}}$ is adequate for our analysis as it provides $\sim$5--10 times better resolution than our per-order Cram{\'e}r-Rao bound, ensuring the grid discretization does not limit our precision. We determined the best-fit RV shift for each order as the velocity minimizing $\chi^2$. This procedure yielded one differential RV measurement per echelle order. We estimate final uncertainties through bootstrap resampling rather than per-order error propagation, as described below.

\begin{figure}[ht]
    \centering
    \begin{overpic}[width=0.48\textwidth]{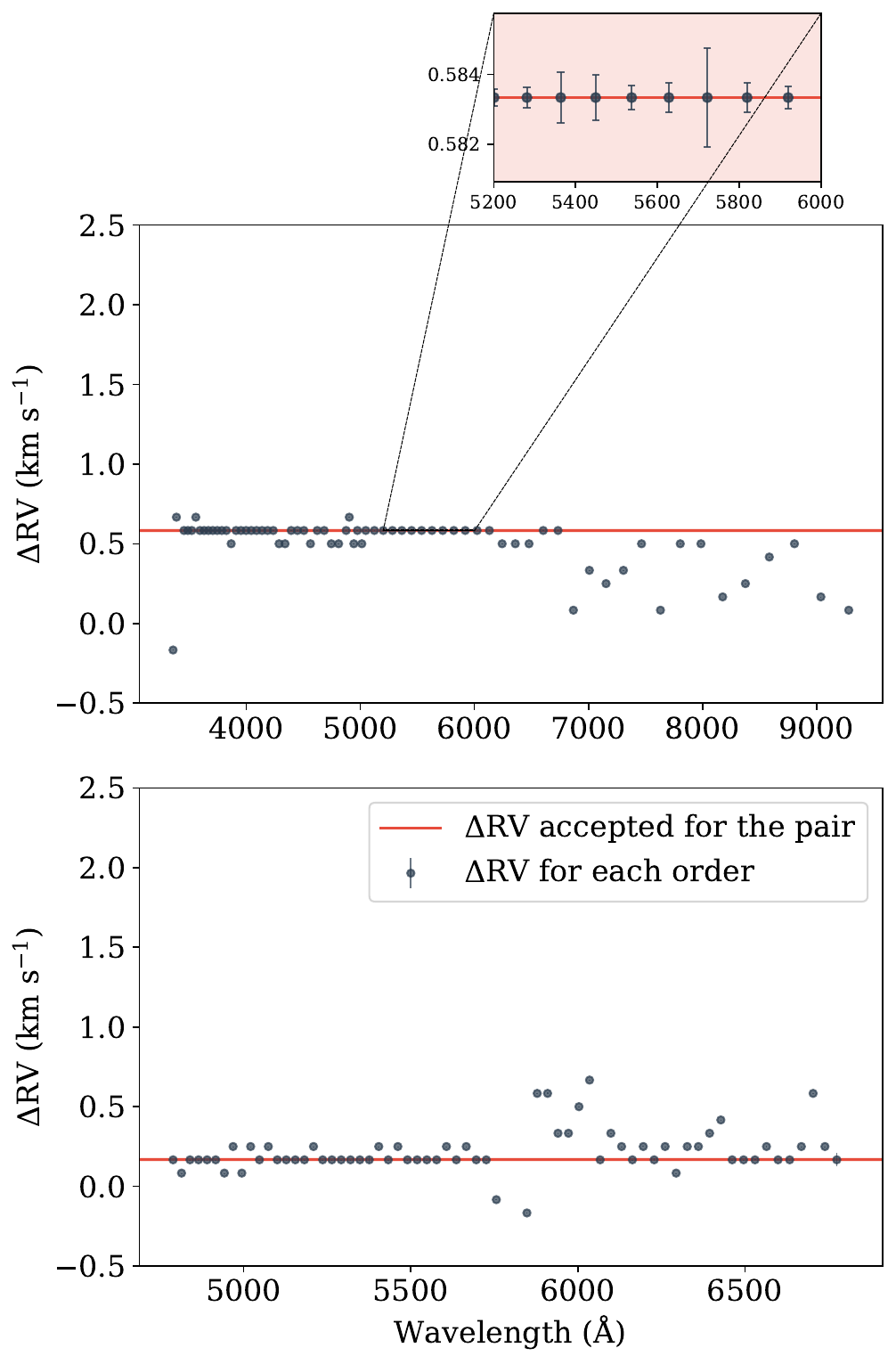}
        \put (27,79) {\makebox(0,0){\footnotesize{Pair 1: Magellan/MIKE}}}
        \put (27,75) {\makebox(0,0){\footnotesize{$\Delta$RV = $583.3\pm9.2~\rm ms^{-1}$}}}
        \put (27,29) {\makebox(0,0){\footnotesize{Pair 2: VLT/UVES}}}
        \put (27,25) {\makebox(0,0){\footnotesize{$\Delta$RV = $166.7\pm8.9~\rm ms^{-1}$}}}
    \end{overpic}
    \caption{Per-order differential RV measurements as a function of wavelength for two representative wide-binary pairs. \textit{Top panel:} Magellan/MIKE observations of the pair Gaia DR3 6473063860275183616 and Gaia DR3 6472230636619614464. The small plot above shows a zoomed in region of this plot.  \textit{Bottom panel:} VLT/UVES observations of the pair Gaia DR3 2919963104221492480 and Gaia DR3 2919963104216198528. Each point represents the $\Delta$RV measured from a single echelle order, with error bars showing the $1\sigma$ uncertainty estimated from the $\chi^2$ curve. The horizontal red line indicates the combined $\Delta$RV obtained from bootstrap resampling of the median, which down-weights outlying orders without requiring sigma clipping.}
    \label{fig:rv_vs_wavelength}
\end{figure}

For bootstrap resampling, we drew  $N$ orders with replacement ($N =$ number of valid orders), compute median of that bootstrap sample, and repeat this for 10,000 iterations. The final differential RV was taken as the median of the bootstrap distribution, while the standard deviation of the distribution gave us an estimate of the uncertainty in differential RV. The differential RV between the components of each binary is then $\Delta \rm{RV} = \rm{RV}_{\rm B} - \mathrm{RV}_{\rm A}$. The high SNR and high-resolution of the C3PO spectra returned per-order RV precisions of $20$--$40~\rm{m\,s^{-1}}$ for most pairs. This improvement over Gaia follows from the Cram{\'e}r-Rao scaling: our spectra have higher resolution (R $\approx$ 50,000 vs 11,500), higher SNR ($\sim$150--300 vs $\approx$20--50), and broader wavelength coverage (3500--9000 \AA~vs 845--872 nm, providing $\sim10\times$ more spectral lines). Combining across $\sim$30--50 valid echelle orders yields final differential RV precisions of $\sim$8--15 $\rm m~s^{-1}$ per binary pair.

To demonstrate the improvement in statistical power, we compared our differential RVs to those derived from \textit{Gaia} for the 84 pairs with available \textit{Gaia} RV measurements. Measured RVs along with their instrument name, \textit{Gaia} ID and \textit{Gaia} proper motion info are given for a few of the sources in Table~\ref{tab:rvdata}. Figure~\ref{fig:rv_vs_wavelength} shows the per-order differential RV measurements as a function of wavelength for two representative binary pairs observed with Magellan/MIKE and VLT/UVES. Comparing our uncertainties to the \textit{Gaia} DR3 RV uncertainties for all 84 pairs with available \textit{Gaia} RVs, we find a median improvement factor of $\xi \approx 24$, meaning our measurements have $\sim$24 times better precision than \textit{Gaia} DR3 RVs. As Fisher information scales as $\sigma^{-2}$, this precision improvement translates directly to increased statistical power for constraining gravity models.

\begin{deluxetable*}{llcr@{$\,\pm\,$}lrr}
\tablecaption{Representative differential radial velocities and \textit{Gaia} astrometry for the wide-binary sample. The full table for all 57 pairs is available in machine-readable form. $\Delta$RV is the differential radial velocity measured in this work; $\Delta\mu_\alpha$ and $\Delta\mu_\delta$ are the differential proper motions from \textit{Gaia} DR3. Excerpt only; the complete table for all pairs, including coordinates, parallaxes, and FLAME masses, is provided in the machine-readable version.\label{tab:rvdata}}
\tablehead{\colhead{\textit{Gaia} DR3 ID (A)} & \colhead{\textit{Gaia} DR3 ID (B)} & \colhead{Instrument} & \multicolumn{2}{c}{$\Delta$RV (m\,s$^{-1}$)} & \colhead{$\Delta\mu_\alpha$ (mas\,yr$^{-1}$)} & \colhead{$\Delta\mu_\delta$ (mas\,yr$^{-1}$)}}
\startdata
6004256802548180736 & 6004256909931771136 & Magellan/MIKE & -590.1 & 1.5 & -2.321 & 1.287 \\
6224633983987511552 & 6224633983987510528 & Magellan/MIKE & -753.3 & 2.7 & -2.783 & 5.366 \\
6666179127640352512 & 6666179333798782976 & Magellan/MIKE & -366.3 & 3.1 & 1.307 & -4.425 \\
2356080043380354816 & 2356290256259997696 & Magellan/MIKE & -568.1 & 4.3 & -1.051 & -0.119 \\
4464207428579939840 & 4464207428577214464 & Magellan/MIKE & -422.7 & 4.6 & 15.219 & -2.497 \\
6463958151650359040 & 6463958151650358784 & Magellan/MIKE & -937.3 & 4.7 & 3.572 & -8.670 \\
5296896205962076416 & 5296896240321814144 & VLT/UVES & 43.6 & 6.1 & -2.484 & 2.483 \\
4940794866807373952 & 4940794488850252928 & VLT/UVES & 357.4 & 6.2 & 1.057 & -0.564 \\
2919963104216198528 & 2919963104221492480 & VLT/UVES & 181.0 & 9.6 & 0.124 & 5.619 \\
5195685713943687168 & 5195685713943683072 & VLT/UVES & -432.6 & 10.4 & -0.572 & -0.416 \\
5863709946382541952 & 5863709946382545280 & VLT/UVES & -220.2 & 13.3 & -0.619 & 0.171 \\
4837828175549438080 & 4837828179846041728 & VLT/UVES & -816.2 & 16.2 & 0.371 & 1.597 \\
\enddata
\end{deluxetable*}

We note that most previous wide binary MOND tests have relied primarily on the sky-projected velocity $v_\perp$ derived from Gaia proper motions rather than Gaia RVs. For nearby pairs in our sample (with typical distances of $\sim$50--150 pc and proper motion uncertainties of $\sim$0.02--0.05 mas\,yr$^{-1}$), the resulting $v_\perp$ uncertainties are typically $\sim$5--20 m\,s$^{-1}$, comparable to our differential RV precision. The primary advantage of our measurements is therefore not a dramatic precision improvement over $v_\perp$, but rather the ability to constrain the full three-dimensional kinematics of each system. This enables hierarchical Bayesian inference of individual orbital elements rather than relying on ensemble statistics of projected velocities.

\begin{figure}
    \centering
        \includegraphics[width=\columnwidth]{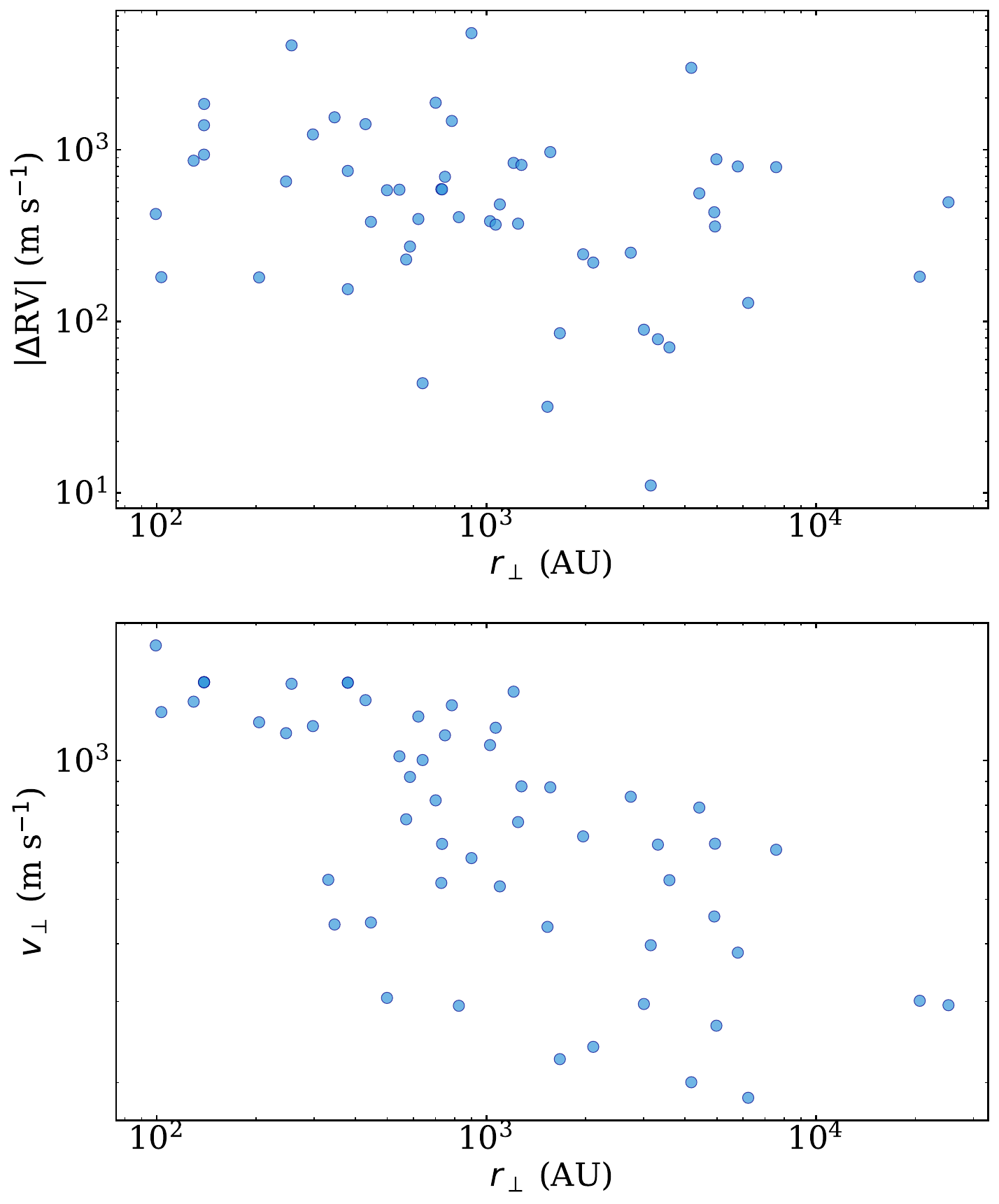}
    \caption{Differential radial velocity $|\Delta{\rm RV}|$ from our high-precision measurements (top) and sky-projected velocity $v_{\perp}$ derived from \textit{Gaia} proper motions (bottom) plotted against the projected separation $r_{\perp}$ for the 57 selected bound systems. The scatter at fixed separation reflects the distribution of orbital phases, eccentricities, and viewing geometries across our sample.
    \label{fig:velocity_rproj}}
\end{figure}

Figure \ref{fig:velocity_rproj} shows the distribution of measured velocity differences as a function of projected separation for the 57 selected bound systems. The top panel displays our differential RV measurements, and the bottom panel shows the sky-projected velocity differences derived from Gaia proper motions. The scatter at fixed separation arises from the distribution of orbital phases, eccentricities, and viewing geometries across our sample.

\subsection{Bound System Selection and Tertiary Contamination}\label{sec:bound}
\label{sec:bound_selection}

To identify gravitationally bound systems among the 85 pairs with
differential RV measurements, we apply a kinematic selection based on the scaled velocity
$\tilde{v}$, following the framework of \cite{Cookson2026}. For each system, we compute
the sky-projected relative velocity $v_{\rm sky}$ from the differential proper motions:
\begin{equation}
v_{\rm sky} = 4.74047 \times d \times \sqrt{(\Delta\mu_\alpha)^2 + (\Delta\mu_\delta)^2}
\label{eq:vsky}
\end{equation}
where $d$ is the mean distance in kpc and the proper motion differences are in
mas\,yr$^{-1}$. We also compute the projected separation $r_{\rm sky}$ and the total
system mass $M_{\rm tot}$ from Gaia FLAME masses \citep{Gaiaflame2}. We then define the
scaled velocity as,
\begin{equation}
\tilde{v} \equiv \frac{v_{\rm sky}}{\sqrt{GM_{\rm tot}/r_{\rm sky}}}
\label{eq:vtilde}
\end{equation}
This dimensionless quantity provides a natural measure of how fast a system is moving
relative to its expected Keplerian velocity. Gravitationally bound systems
orbits have $\tilde{v} \lesssim 2$ for the vast majority of orbital configurations, while
chance alignments and unbound pairs produce systematically higher values of $\tilde{v}$.

We select systems satisfying $\tilde{v} < 2.5$. This threshold is motivated by the
analysis of \cite{Cookson2026}, who showed that chance alignments and contamination from
unresolved close binaries can mimic anomalous velocity signals, and that a cut on
$\tilde{v}$ provides an effective criterion for removing such contaminants. Of the 85
systems, 57 satisfy this criterion and are retained for our MOND analysis.

As an additional verification, Figure~\ref{fig:velocity_rproj} shows the distribution of
measured velocity differences as a function of projected separation for the 57 selected
bound systems. The top panel displays our differential RV measurements, and the bottom
panel shows the sky-projected velocity differences derived from \textit{Gaia} proper
motions. The velocities of these 57 systems are consistent with expectations for
gravitationally bound systems: the scatter at fixed separation arises naturally
from the distribution of orbital phases, eccentricities, and viewing geometries, and no
system exhibits anomalously high velocities that would suggest chance alignment.

A concern for any wide binary study is contamination from unresolved tertiary companions,
which would inflate the observed velocity difference and bias inferred parameters. The
C3PO sample selection provides several safeguards against this contamination.

First, the absolute magnitude constraint ($\Delta M_G < 1$ mag)  {and the color similarity requirement (} $|\Delta(\rm BP-RP)| < 0.15$  {mag) act together. Neither is sufficient on its own. An unresolved inner binary of two equal stars is only 0.75 mag brighter than either star alone, so a system in which one component is such a pair passes the magnitude criterion. What excludes it is the color requirement}. {Two lower-mass stars that together match the luminosity of a single star are each redder than that star, so the composite is redder than the companion of the same magnitude and fails the color cut. For the F, G and K dwarfs in this sample the required mass difference is about 20 percent, which corresponds to a color offset well above the 0.15 mag tolerance. The two criteria therefore exclude luminous tertiaries only when applied jointly, which is how the C3PO sample was constructed \mbox{\citep{C3POI}}\hskip0pt.} 

 {Second}, the C3PO spectra underwent visual inspection to exclude obvious double-lined
spectroscopic binaries (SB2s) \citep{C3POI}. While this does not exclude all
spectroscopic binaries, it removes systems with companions bright enough to produce
detectable spectral features.

We acknowledge that low-mass, faint tertiary companions cannot be completely excluded. As an order-of-magnitude estimate, a companion with mass $M_3 \lesssim 0.3\,{\rm M}_\odot$ at an orbital period of a few months produces a radial velocity kick of $K \sim 50$--$150\,\mathrm{m\,s^{-1}}$ on its host star. This is 3--10 times our per-pair measurement uncertainty and would produce outliers in our 4D joint fit (RV $+$ proper motions). We estimate that the maximum contamination fraction consistent with the observed velocity--separation trend is $\lesssim 5\%$, negligible at our current sample size. Importantly, triple contamination biases observed velocities upward---opposite to the direction of the MOND signal---so any residual contamination would not drive our result toward lower $a_0$. We also note that our bound selection criterion preferentially removes systems with
anomalously high velocities, which could arise from tertiary contamination.
\section{MOND orbits and geometry} \label{sec:formulation}
\subsection{MOND formulation}

According to Milgrom's original MOND formulation \citep{Milgrom1983}, gravitational acceleration is modified when it falls below a scale of $a_0 \approx 1.2 \times 10^{-10}$ m s$^{-2}$. This modification is implemented through an interpolating function that transitions smoothly between the Newtonian and MOND regimes. We consider a generalized interpolating function of the form,
\begin{equation}
\mu(x) = \frac{x}{(1 + x^b)^{1/b}}
\label{eq:mu_general}
\end{equation}
where $x = a/a_0$ is the acceleration in units of $a_0$, and $b$ is a sharpness parameter controlling the transition. Two commonly used forms are:

\begin{itemize}
\item $b = 1$: the ``simple'' interpolating function, $\mu(x) = x/(1+x)$
\item $b = 2$: the ``standard'' interpolating function, $\mu(x) = x/\sqrt{1+x^2}$
\end{itemize}

In both cases, $\mu(x) \rightarrow 1$ for $x \gg 1$ (Newtonian regime) and $\mu(x) \rightarrow x$ for $x \ll 1$ (deep MOND regime). It is to note that, the exact shape of the interpolating function is not fixed by first principle, but is instead chosen to satisfy the limiting behavior. Many studies have adopted the simple form ($b=1$) \citep{Famaey2005, McGaugh2011MOND}. The standard form ($b=2$) is also used, as it was originally introduced by Milgrom \citep{Milgrom1983, Sanders2002}. Beyond these, alternative interpolating functions have also been proposed to better match external-field constraints or specific galactic datasets \citep{Zhao2006, Gentile2011}. For our study, we chose both $b=1$ and $b=2$, and we compare our final results between them.

\begin{figure}
\centering
\includegraphics[width=\columnwidth]{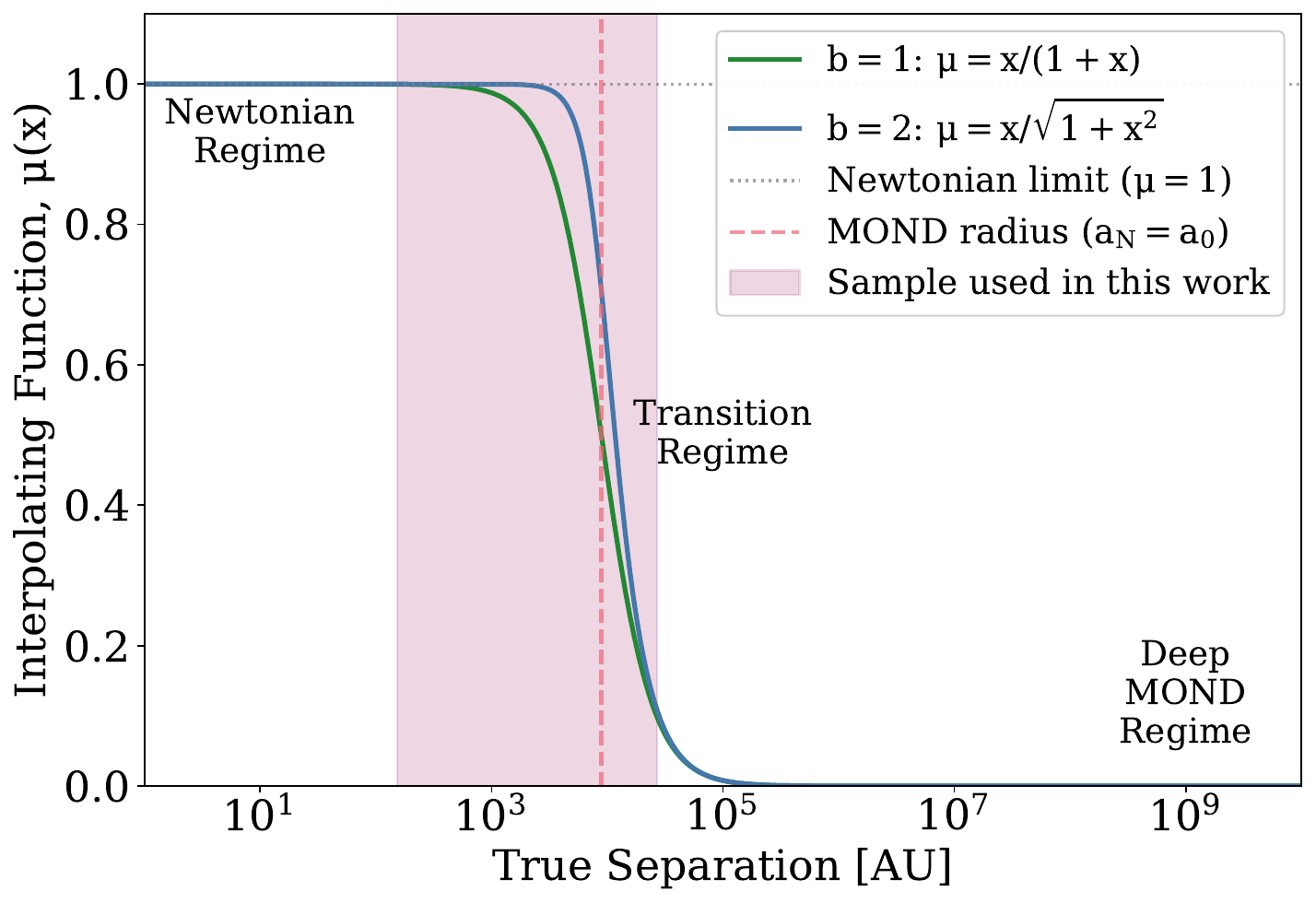}
\caption{Transition from Newtonian to MOND regime for two different interpolating functions ($b=1$ is green
and $b=2$ is blue).
The pink shaded region highlights the range of true separations spanned by our wide binary sample.
The MOND radius ($a_N = a_0$) is marked with red dashed line and the Newtonian limit is marked as gray dotted
lines.
The transition is sharper for $b=1$ than $b=2$.
\label{fig:interp}}
\end{figure}

For wide binary systems like the C3PO sample, we must also account for the external gravitational field ($a_{\rm ext}$) from the galaxy. Unlike Newtonian gravity, where a uniform external field produces only tidal effects that are negligible for bound systems, in MOND, the external field modifies the internal dynamics even when spatially uniform. This is because the MOND acceleration depends on the total acceleration, not just the internal component. Because these systems are located in the solar neighborhood at galactocentric radius $R_{\rm GC} \approx 8$~kpc, where the galactic gravitational field produces $a_{\rm ext} \approx 2 \times 10^{-10}$~m\,s$^{-2}$, the external acceleration is comparable to $a_0$ and cannot be neglected. The total acceleration for each star depends on both the internal acceleration ($a_{\rm int}$) from the companion and the external acceleration ($a_{\rm ext}$) from the galaxy. The angle ($\theta$) between these vectors varies as the stars orbit each other. The magnitude of the total acceleration is,

\begin{equation}
a_{\rm tot} = \sqrt{a_{\rm int}^2 + a_{\rm ext}^2 + 2 a_{\rm int} a_{\rm ext} \cos\theta}
\label{eq:atot}
\end{equation}

The effective acceleration in MOND is then,
\begin{equation}
\begin{aligned}
a_{\rm eff} = \Big(
    &\mu_{\rm tot}^2\, a_{\rm int}^2 
     + (\mu_{\rm tot}-\mu_{\rm ext})^2\, a_{\rm ext}^2  \\
    &+ 2\,\mu_{\rm tot}(\mu_{\rm tot}-\mu_{\rm ext})
        a_{\rm int} a_{\rm ext} \cos\theta
\Big)^{1/2}
\end{aligned}
\label{eq:aeff}
\end{equation}
where $\mu_{\rm tot} = \mu(a_{\rm tot}/a_0)$ and $\mu_{\rm ext} = \mu(a_{\rm ext}/a_0)$.

For our calculations, we adopt an external acceleration value of $a_{\rm ext} = 2.1 \times 10^{-10}~\rm{m\,s^{-2}}$  for the solar neighborhood, with the external field pointing toward the galactic center at $(\alpha_{\rm GC}, \delta_{\rm GC}) = (266.4051^\circ, -28.936175^\circ)$ \citep{Desmond2024, Oort1960}.

Figure~\ref{fig:interp} compares the two interpolating functions and how they change throughout different separations between the two components of a binary pair. It shows that the predicted true separation for our sample lies in the transition regime between Newtonian and MOND. Figure~\ref{fig:posterior_b12} directly compares the two posteriors. The $b=2$ posterior is systematically shifted toward higher $a_0$ values compared to $b=1$. It shows that constraints on $a_0$ from wide-binary dynamics depend on the assumed form of the interpolating function.

In addition to the full MOND formulation with the external field effect, we also perform a supplementary test using an effective gravity boost factor $\gamma$, in which the gravitational interaction is simply rescaled as $G_{\rm eff} = \gamma G_{\rm N}$. This alternative formulation does not assume any particular MOND interpolating function or external field treatment, and allows direct comparison with several previous wide-binary studies that have reported results in terms of $\gamma$ \citep{Hernandez2024a, Chae2023, Chae2024b, Chae2026, Banik2024}. Because the C3PO systems reside primarily in the MOND transition regime, where the MOND predicted gravity boost varies, the $\gamma$ parameterization is less natural for this dataset than the direct inference of $a_0$. We therefore treat the $\gamma$ test as a supplementary analysis and present the full details in Appendix~\ref{appendix:gamma_test}.

\subsection{Orbital elements and Keplerian motion}\label{subsec:orbital}
We describe each wide binary pair with the six Keplerian orbital elements: semi-major axis $a$, eccentricity
$e$, inclination $i$, longitude of ascending node $\Omega$, argument of periastron $\omega$, and mean anomaly
$M$. Given a value of $M$, we solve Kepler's equation,
\begin{equation}
M = E - e\sin E
\label{eq:MtoE}
\end{equation}
for the eccentric anomaly $E$ using the Newton--Raphson method with 7 iterations, which is sufficient for convergence at the eccentricities in our sample. The instantaneous separation between the two stars is then computed directly from $E$ as,
\begin{equation}
r = a(1-e\cos E).
\label{eq:r}
\end{equation}
We note that in some recent wide-binary studies \citep{Chae2026, Hernandez2024a}, the separation is instead expressed in terms of the true anomaly $\nu$ via $r=a(1-e^2)/(1+e\cos\nu)$, where $\nu$ is obtained from $E$ through the relation $\nu=2\tan^{-1}[\sqrt{(1+e)/(1-e)}\tan(E/2)]$. However, since we already solve for $E$ from Kepler's equation, the separation and velocity can be computed directly in terms of $E$ without the intermediate conversion to $\nu$, resulting in a simpler formulation.

\subsection{MOND-modified velocities}\label{subsec:mondvel}
The key difference between MOND and Newtonian dynamics lies in the orbital velocity.
In Newtonian mechanics, the velocity magnitude can be determined by using $a_N = G(M_1+M_2)/r^2$.
In MOND, we replace this with the effective acceleration $a_{\rm eff}$ from Eq.~\eqref{eq:aeff}.
Using the eccentric anomaly, the position of star B relative to star A in the orbital plane is,
\begin{subequations}
\begin{align}
x_{\rm orb} &= a(\cos E-e), \\
y_{\rm orb} &= a\sqrt{1-e^2}\sin E.
\end{align}
\end{subequations}
The velocity components in the orbital plane follow directly from differentiation with respect to time. Using the relation $\dot{E}=n/(1-e\cos E)$, where $n=\sqrt{G(M_1+M_2)/a^3}$ is the mean motion, the Newtonian orbital velocity components are,
\begin{subequations}
\begin{align}
\dot{x}_{\rm orb} &= -\frac{na}{1-e\cos E}\sin E, \\
\dot{y}_{\rm orb} &= \frac{na\sqrt{1-e^2}}{1-e\cos E}\cos E.
\end{align}
\end{subequations}
In MOND, we scale the velocity magnitude by the ratio of MOND to Newtonian acceleration. Defining
\begin{equation}
\beta = \sqrt{a_{\rm eff}/a_N},
\label{eq:vmag}
\end{equation}
where $a_N=G(M_1+M_2)/r^2$ and $a_{\rm eff}$ is from Eq.~\eqref{eq:aeff}, the MOND-modified velocity components become $v_x=\beta\dot{x}_{\rm orb}$ and $v_y=\beta\dot{y}_{\rm orb}$. For each star in the binary system, the velocities in the orbital frame are,
\begin{subequations}
\begin{equation}
\vec{v}_{1,{\rm orb}} = \frac{M_2}{M_1+M_2}
\begin{pmatrix} v_x \\ v_y \\ 0 \end{pmatrix}
\label{eq:v1orb}
\end{equation}
\begin{equation}
\vec{v}_{2,{\rm orb}} = -\frac{M_1}{M_1+M_2}
\begin{pmatrix} v_x \\ v_y \\ 0 \end{pmatrix}
\label{eq:v2orb}
\end{equation}
\end{subequations}

\subsection{Coordinate transformation}
To convert orbital velocities to observed velocity components, we transform the coordinate from the orbital plane to International Celestial Reference System (ICRS). The rotation matrix is,
\begin{equation}
\mathcal{R} = \mathcal{R}_z(\Omega) \cdot \mathcal{R}_x(i) \cdot \mathcal{R}_z(\omega),
\label{eq:rotation}
\end{equation}
where $\mathcal{R}_z$ and $\mathcal{R}_x$ are rotation matrices about the $z$ and $x$ axes, respectively. The position and velocity vectors in ICRS coordinates are then,
\begin{subequations}
\begin{align}
\vec{r}_{\rm ICRS} &= \mathcal{R} \vec{r}_{\rm orb} \label{eq:rICRS}\\
\vec{v}_{i,\rm ICRS} &= \mathcal{R} \vec{v}_{i,\rm orb}, \quad i \in \{1,2\} \label{eq:vICRS}
\end{align}
\end{subequations}
where $\vec{r}_{\rm orb}=(x_{\rm orb},y_{\rm orb},0)^T$.

The angle $\theta$ in the Eq.~\eqref{eq:aeff} is computed as,
\begin{equation}\cos\theta = \hat{r}_{\rm ICRS} \cdot \hat{a}_{\rm ext}\label{eq:theta}\end{equation}

\noindent where $\hat{r}_{\rm ICRS}$ is the unit vector pointing from star A to star B, and $\hat{a}_{\rm ext}$ points toward the Galactic center. This angle varies throughout the orbit, creating time-dependent modifications to the dynamics.

\subsection{Observable quantities}

Because our high-precision RV measurements are differential, we formulate all observables in terms of relative quantities. This differential approach is central to our method: it cancels many systematic effects common to both stars, including instrumental zero-point offsets and barycentric correction errors. For proper motions, we similarly use the differential values $\Delta\mu_\alpha$ and $\Delta\mu_\delta$, primarily for consistency with our orbital model.

\begin{itemize}
    \item \textit{Projected Separation}: For each star in a system, we define a local celestial coordinate system with unit vectors $\hat{r}$ (radial), $\hat{e}$ (east), and $\hat{n}$ (north) based on the star's right ascension $\alpha$ and declination $\delta$:
\begin{subequations}
\begin{align}
\hat{r} &= (\cos\delta\cos\alpha, \cos\delta\sin\alpha, \sin\delta) \\
\hat{e} &= (-\sin\alpha, \cos\alpha, 0) \\
\hat{n} &= (-\sin\delta\cos\alpha, -\sin\delta\sin\alpha, \cos\delta)
\end{align}
\end{subequations}

The projected separation on the sky computed using the coordinates of star A is,
\begin{equation}
s_{\rm proj} = \sqrt{r_E^2 + r_N^2}
\label{eq:sproj}
\end{equation}
where $r_E = \vec{r}_{\rm ICRS} \cdot \hat{e}_A$ and $r_N = \vec{r}_{\rm ICRS} \cdot \hat{n}_A$.

\item 
\textit{Differential Radial Velocity}: Each binary system has a systematic velocity $\vec{v}_{\rm sys}$ common to both the components. The RV for each star is, 
\begin{subequations}
\begin{align}
v_{r,1} &= (\vec{v}_{1,\rm ICRS} + \vec{v}_{\rm sys}) \cdot \hat{r}_A \label{eq:vr1}\\
v_{r,2} &= (\vec{v}_{2,\rm ICRS} + \vec{v}_{\rm sys}) \cdot \hat{r}_B \label{eq:vr2}
\end{align}
\end{subequations}

The differential RV is,
\begin{equation}
\Delta v_r = v_{r,2} - v_{r,1}
\label{eq:deltavr}
\end{equation}

\item \textit{Differential Proper Motion}: The tangential velocity components are,
\begin{align}
\vec{v}_{i,t} &= \vec{v}_{i,\rm ICRS} + \vec{v}_{\rm sys} - [(\vec{v}_{i,\rm ICRS} + \vec{v}_{\rm sys}) \cdot \hat{r}_i]\hat{r}_i
\end{align}
where $i \in \{1,2\}$. The proper motions are,
\begin{subequations}
\begin{align}
\mu_{\alpha,i} &= -\frac{\vec{v}_{i,t} \cdot \hat{e}_i}{4.74047 \times d_i} \label{eq:pmra}\\
\mu_{\delta,i} &= \frac{\vec{v}_{i,t} \cdot \hat{n}_i}{4.74047 \times d_i} \label{eq:pmdec}
\end{align}
\end{subequations}
where $d_i$ is the distance in parsecs and the factor 4.74047 converts $\rm{km\,s^{-1}\,pc^{-1}}$ to mas yr$^{-1}$. The differential proper motions are,
\begin{subequations}
\begin{align}
\Delta\mu_\alpha &= \mu_{\alpha,2} - \mu_{\alpha,1} \label{eq:dpmra}\\
\Delta\mu_\delta &= \mu_{\delta,2} - \mu_{\delta,1} \label{eq:dpmdec}
\end{align}
\end{subequations}

\end{itemize}

\begin{figure*}
    \centering
    \scalebox{0.52}{\begin{tikzpicture}[scale=3.5]

\tikzset{
    every node/.style={font=\Large}, %
}

\filldraw[draw=none,fill=gray!20] (-2,0.7) -- (-1.2,-1.2) -- (2.8,-1.2) -- (2,0.7) -- cycle;

\definecolor{blue}{HTML}{0072B2}
\definecolor{red}{HTML}{B00020}
\definecolor{orange}{HTML}{E69F00}
\definecolor{teal}{HTML}{009E73}

\tikzstyle{vec}=[->, thick]
\tikzstyle{orb}=[thick, black]
\tikzstyle{starA}=[circle, fill=yellow!70, draw=black]
\tikzstyle{starB}=[circle, fill=orange!80, draw=black]

\draw[blue, ->, thick] (0,0) -- (0.4,-0.9) node[left] {$x(+\alpha)$};
\draw[blue, ->, thick] (0,0) -- (1.7,0) node[above] {$y(+\delta)$};
\draw[blue, ->, thick] (0,0) -- (0,1.7) node[right] {$z \; (\rm Away \; from \; Observer)$};

\draw[->, orb, rotate=130, thick, black] (-0.5,0) ellipse (1.6 and 0.9);

\draw[black, thin, rotate=130] (0,-0.85) -- (0,0.85);
\draw[black, thin, rotate=130] (-2.1,0) -- (1.1,0);
\draw[red, thick] (0,0) -- (0.2,0.85);
\draw[black, thin] (-0.6525,-0.55) -- (-1.2,-0.3);

\filldraw[starA] (0.2,0.85) circle (0.05);
\filldraw[starA] (0,0) circle (0.05);

\draw[->, red, thick] (0.14,0) arc (0:-140:0.15);
\draw[->, red, thick] (-0.18,-0.16) arc (-140:-205:0.45);
\draw[->, red, thick] (-0.25, 0.31) arc (120:75:0.45);
\draw[->, red, thick] (-0.9, -0.44) arc (150:126:0.45);

\node[thick, red] at (-0.17,0.45) {$\nu$};
\node[thick, red] at (-0.4,0) {$\omega$};
\node[thick, red] at (-0.05,-0.25) {$\Omega$};
\node[thick, red] at (-0.9,-0.3) {$i$};
\node[thin, black] at (2.2,-0.8) {Sky Plane};
\node[thin, black] at (-0.5,-1.6) {To Observer};
\node[thick, red, rotate=80] at (0.1,0.8) {$r$};
\node[thick, black] at (0.65,1.06) {Star B};
\node[thick, black] at (0.64,0.13) {Star A};
\node[thick,black] at (-1, -1.1) {Ascending Node};

\draw[->, thick, black]
  (-1,-1)
  .. controls (-0.95,-0.74) .. (-0.85,-0.7)
  .. controls (-0.75,-0.66) .. (-0.68,-0.58);

\draw[->, thick, black]
  (0.4, 0.1)
  .. controls (0.2,0.1) .. (0.05, 0.02);

\draw[->, thick, black]
  (0.4, 1.05)
  .. controls (0.25,1) .. (0.22,0.9);

\draw[->, double] (0,-1.5) -- (0,-1.7);

\end{tikzpicture}}
    \hspace{1cm}
    \scalebox{0.55}{\begin{tikzpicture}[scale=3.5]

\tikzset{
    every node/.style={font=\Large}, %
}

\definecolor{blue}{HTML}{0072B2}
\definecolor{red}{HTML}{B00020}
\definecolor{orange}{HTML}{E69F00}
\definecolor{teal}{HTML}{009E73}

\tikzstyle{vec}=[->, thick]
\tikzstyle{orb}=[thick, black]
\tikzstyle{starA}=[circle, fill=yellow!70, draw=black]
\tikzstyle{starB}=[circle, fill=yellow!70, draw=black]

\draw[->, orb, thick, black] (0,0) ellipse (1.6 and 0.9);
\draw[dashed, orb, thin, black!20] (0,0) ellipse (1.6 and 1.6);

\draw[black, thin] (0,-1.6) -- (0,1.6);
\draw[black, thin] (-1.6,0) -- (1.6,0);
\draw[blue, thin] (1,0) -- (1.3,0.53);
\draw[blue, thin] (1,0) -- (1.3,0.53);
\draw[blue, thin] (1.3,0) -- (1.3,0.93);
\draw[blue, thin] (0,0) -- (1.3,0.93);
\draw[->, red, thick] (0.4,-0.1) -- (0,-0.1);
\draw[->, red, thick] (0.6,-0.1) -- (1,-0.1);

\filldraw[starA] (1.3,0.53) circle (0.05);
\filldraw[starB] (1,0) circle (0.05);

\draw[->, red, thick] (0.3,0) arc (0:36:0.3);
\draw[->, red, thick] (1.15,0) arc (0:60:0.15);

\node[thick, red] at (0.5,-0.1) {$ae$};
\node[thick, red] at (0.37,0.12) {$E$};
\node[thick, red] at (1.2,0.11) {$\nu$};
\node[thick, red] at (1.15,0.41) {$r$};
\node[thick, black] at (0.5,-1.07) {Auxiliary Circle};
\node[thick, black] at (-0.75,-1.07) {Elliptical Orbit};
\node[thick, black] at (1.95,0.68) {Star B};
\node[thick, black] at (1.23,-0.3) {Star A};

\draw[->, thick, black]
  (-0.8,-1)
  .. controls (-1.05,-0.84) .. (-0.95,-0.74);
\draw[->, thick, black]
  (0.55, -1.15)
  .. controls (0.5,-1.3) .. (0.85,-1.35);

\draw[->, thick, black]
  (1.7, 0.68)
  .. controls (1.4,0.65) .. (1.35,0.58);

\draw[->, thick, black]
  (1.25, -0.25)
  .. controls (1.2,-0.1) .. (1.05,-0.05);

\end{tikzpicture}}\\[0.3cm]
    \caption{Relative orbits and Keplerian elements for wide binary systems. \textit{Left:} Three-dimensional view showing the relative orbit of star B around star A (both yellow circles). The coordinate system is defined with the $z$-axis pointing away from the observer (RV direction), and $x$ and $y$ axes aligned with right ascension ($\alpha$) and declination ($\delta$). The orbital orientation is specified by inclination $i$ (angle between orbital plane and sky plane), longitude of ascending node $\Omega$ (where the orbit crosses the sky plane moving away from the observer), argument of periastron $\omega$ (angle from ascending node to periastron), and true anomaly $\nu$ (angle from periastron to current position). The instantaneous separation between stars is $r$. \textit{Right:} Orbital plane view of the same diagram showing the relationship between eccentric anomaly $E$, true anomaly $\nu$, and the auxiliary circle. The distance $ae$ represents the offset of the focus from the ellipse center, where $a$ is the semi-major axis and $e$ is the eccentricity.}
    \label{fig:orbit_diagram}
\end{figure*}

Figure~\ref{fig:orbit_diagram} illustrates the relative orbit in both three-dimensional view (in left) and projection onto the orbital plane (in right), with the key orbital elements indicated.

\section{Model \& Inferences} \label{sec: Methods}

\begin{figure*}
    \centering
    {\usetikzlibrary{fit,positioning}

\begin{tikzpicture}[
    scale=1.3,
    shorten >=1pt,->,draw=black!70,
    neuron/.style={circle,minimum size=40,inner sep=0},
    param/.style={neuron,fill=cyan}, %
    inter/.style={neuron,fill=black!40}, %
    gauss/.style={neuron,fill=red}, %
    obs/.style={neuron,fill=yellow}, %
    beta/.style={neuron,fill=yellow}, %
    box/.style={rectangle,draw=black,rounded corners,thick,inner sep=10pt}
]

\node[param] (a0) at (-0.75,5.5) {$\log_{10}a_0$};
\node[gauss] (ua0) at (0.75,5.5) {$\mathcal{U}$};

\node[param] (a)   at (0,4) {$a$};
\node[param] (alpha)   at (-1.5,4) {$\alpha$};
\node[inter] (nu)  at (-3,1) {$\nu$};
\node[param] (M)   at (-3,4) {$M$};
\node[param] (i)   at (3,4) {$\cos i$};
\node[param] (omega)   at (4.5,4) {$\omega$};
\node[param] (Omega)   at (4.5,2.5) {$\Omega$};

\node[param] (M1M2) at (1.5,4)
  {\parbox{1cm}{\centering $M_1$ \\ $M_2$}};

\node[inter] (e) at (-1.5,1) {$e$};
\node[inter] (rtrue) at (-1.5,-0.5) {$r_{\rm true}$};
\node[inter] (theta) at (2.25,1) {$\theta_{\rm ext}$};
\node[inter] (R) at (3.75,1) {$R$};
\node[inter] (vproj) at (3.75,-0.5) {$\parbox{1cm}{\centering $v_{1,\rm proj}$ \\ $v_{2,\rm proj}$}$};

\node[inter] (v1v2) at (0.75,1) {$\parbox{1cm}{\centering $v_1$ \\ $v_2$}$};
\node[inter] (rv) at (0.75,-2) {$\Delta \rm RV$};
\node[inter] (pmra) at (2.25,-2) {$\Delta \mu_{\alpha}$};
\node[inter] (pmdec) at (3.75,-2) {$\Delta \mu_{\delta}$};
\node[inter] (rproj)   at (-1.5,-2) {$r_{\rm proj}$};

\node[gauss] (uM)   at (-3,2.5) {$\mathcal{U}$};
\node[gauss] (ue)   at (-1.5,2.5) {$\mathcal{U}$};
\node[gauss] (ua)   at (0,2.5) {$\mathcal{U}$};
\node[gauss] (gM1M2)   at (1.5,2.5) {$\mathcal{N}$};
\node[gauss] (uR)   at (3,2.5) {$\mathcal{U}_{3}$};
\node[gauss] (gv1v2) at (2.25,-3.5) {$\mathcal{N}_{4}$};

\node[obs] (robs) at (-0.75,-5) {$r_{\rm proj}$};
\node[obs] (rvobs) at (0.75,-5) {$\Delta \rm RV$};
\node[obs] (pmraobs) at (2.25,-5) {$\Delta \mu_{\alpha}$};
\node[obs] (pmdecobs) at (3.75,-5) {$\Delta \mu_{\delta}$};

\draw[->] (i) -- (uR);
\draw[->] (omega) -- (uR);
\draw[->] (Omega) -- (uR);
\draw[->] (uR) -- (R);
\draw[->] (R) -- (rproj);
\draw[->] (R) -- (theta);
\draw[->] (M) -- (uM);
\draw[->] (uM) -- (nu);

\draw[->] (a) -- (ua);
\draw[->] (ua) -- (rtrue);
\draw[->] (e) -- (rtrue);
\draw[->] (alpha) -- (ue);
\draw[->] (ue) -- (e);
\draw[->] (nu) -- (rtrue);

\draw[->] (M1M2) -- (gM1M2);
\draw[->] (gM1M2) -- (v1v2);
\draw[->] (rtrue) -- (rproj);

\draw[->] (a0) -- (ua0);
\draw[->] (ua0) -- (v1v2);
\draw[->] (theta) -- (v1v2);
\draw[->] (rtrue) -- (v1v2);

\draw[->] (v1v2) -- (vproj);
\draw[->] (R) -- (vproj);
\draw[->] (vproj) -- (rv);
\draw[->] (vproj) -- (pmra);
\draw[->] (vproj) -- (pmdec);

\draw[->] (rproj) -- (gv1v2);
\draw[->] (gv1v2) -- (robs);
\draw[->] (rv) -- (gv1v2);
\draw[->] (gv1v2) -- (rvobs);
\draw[->] (pmra) -- (gv1v2);
\draw[->] (gv1v2) -- (pmraobs);
\draw[->] (pmdec) -- (gv1v2);
\draw[->] (gv1v2) -- (pmdecobs);

\node[box, fit=(a)(e)(M)(nu)
             (M1M2)(rtrue)(theta)(i)(omega)(Omega)
             (v1v2)(rv)(pmra)(pmdec)
             (gv1v2)(robs)(rvobs)(pmraobs)(pmdecobs)] (bigbox) {};

\node[anchor=south west,font=\small] at ([xshift=10pt,yshift=10pt]bigbox.south west) {For each pair};

\node[anchor=east] at ([xshift=-0.5cm]bigbox.west) {
    \begin{tikzpicture}[every node/.style={anchor=west}, x=1cm, y=0.6cm]
        \node[param,label=right:{Parameter}] at (3,6) {};
        \node[inter,label=right:{Intermdeiate Variables}] at (3,3) {};
        \node[gauss,label=right:{Distribution}] at (3,0) {};
        \node[obs,label=right:{Observed Variables}] at (3,-3) {};
    \end{tikzpicture}
};

\end{tikzpicture}}
    \caption{Graphical representation of the hierarchical Bayesian model for inferring MOND parameter $a_0$ and orbital elements. The global MOND acceleration scale $\log_{10}a_0$ (blue) is shared across all binary systems and inferred jointly from the full dataset. The box denotes plate notation, indicating that the enclosed structure is replicated for each of the $N=$ 57 binary systems. For each system, we infer six orbital elements: semi-major axis $a$, eccentricity shape parameter $\alpha$ (which determines $e$ via a Beta distribution), mean anomaly $M$, inclination $i$, argument of periastron $\omega$, and longitude of ascending node $\Omega$, along with stellar masses $M_1$ and $M_2$. These parameters determine intermediate quantities (gray): eccentric anomaly $E$, instantaneous separation $r_{\rm true}$, external field angle $\theta_{\rm ext}$, rotation matrix $R$, and stellar velocities $v_1$, $v_2$. The model predicts four observables (yellow): projected separation $r_{\perp}$, differential radial velocity $\Delta v_r$, and differential proper motions $\Delta\mu_\alpha$ and $\Delta\mu_\delta$. Red nodes indicate probability distributions linking parameters to intermediate quantities or observations. $N_4$ denotes a 4-dimensional Gaussian likelihood over the observables. This model structure applies to both interpolating functions tested: $b=1$ (simple) and $b=2$ (standard), where $b$ controls the sharpness of the transition between Newtonian and MOND regimes in Eq.~\eqref{eq:mu_general}.}
    \label{fig:mond_diagram}
\end{figure*}

\begin{table*}
\caption{Prior and likelihood distributions used for {the hierarchical Bayesian} MOND analysis. The global MOND parameter $a_0$ applies to all binary systems. For each system $j$, we infer all the other parameters. Stellar masses $M_{i,j}$ use Gaia FLAME estimates \citep{Gaiaflame} as Gaussian priors. {The semi-major axis prior is the broken power law of Eq.~\eqref{eq:bpl_prior} \citep{Pittordis2018}, sampled through $\eta_j = \ln(a_j/r_\perp)$.} The eccentricity distribution depends on separation-dependent shape parameters $\alpha_k$ for different projected separation bins following \cite{Hwang2022eccentricity}. Separation uncertainty $\sigma_{s,j}$ includes a 5\% floor; $\sigma_{\Delta v_r,j}$ is from our differential RV analysis; $\sigma_{\mu,j}$ is from \textit{Gaia}. We treat the \textit{Gaia} proper motion measurements of the two stars as independent, which is appropriate given their wide angular separations.}
\label{tab:prior}
\centering
\footnotesize
\resizebox{\textwidth}{!}{%
\begin{tabular}{lcl}
\hline\hline
Parameter & Definition & Description \\
\hline
\multicolumn{3}{c}{\textit{Priors}} \\
\hline
$\log_{10} a_0$ & $\mathcal{U}(-13, -7)$,  $\mathcal{U}(-12, -8)$, and $\mathcal{U}(-11, -9)$ & Global MOND acceleration scale in log-space \\
$M_{i,j}$ (M$_{\odot}$) & $\mathcal{N}(M_{{\rm FLAME},i,j},\, \sigma_{{\rm FLAME},i,j}^2)$ & Stellar masses from Gaia FLAME isochrone fitting \\
$\eta_j = \ln(a_j/r_\perp)$ & {$\mathcal{U}(-1,\,2.5)$} & {Semi-major axis log-ratio, flat in $\ln a$ (\"{O}pik's law)} \\
{$a_j$ (population)} & {$\propto (a_j/5\,{\rm kAU})^{-0.6}$ above 5 kAU} & {Broken power-law prior, Eq.~\eqref{eq:bpl_prior} \citep{Pittordis2018}} \\
$\alpha_{k}$ & $\mathcal{N}(\mu_{\alpha,k}, \sigma_{\alpha,k}^2)$ & Eccentricity shape parameter per bin \\
$\cos i_{j}$ & $\mathcal{U}(-1, 1)$ & Inclination \\
$\omega_{j}$ & $\mathcal{U}(0, 2\pi)$ & Argument of periastron \\
$\Omega_{j}$ & $\mathcal{U}(0, 2\pi)$ & Longitude of ascending node \\
$M_{j}$ & $\mathcal{U}(0, 2\pi)$ & Mean anomaly \\
\hline
\multicolumn{3}{c}{\textit{Intermediate Variables}} \\
\hline
$e_j$ & Eq.~\eqref{eq:prior_e} & Eccentricity (from shape parameter) \\
$E_j$ & Eq.~\eqref{eq:MtoE} & Eccentric anomaly \\
$r_{{\rm true},j}$ & Eq.~\eqref{eq:r} & Instantaneous separation \\
$R_j$ & Eq.~\eqref{eq:rotation} & Rotation matrix \\
$\theta_{{\rm ext},j}$ & Eq.~\eqref{eq:theta} & External field angle \\
$v_{1,j}, v_{2,j}$ & Eqs.~(\ref{eq:v1orb}--\ref{eq:v2orb}) & Stellar velocities (orbital frame) \\
$v_{1,{\rm proj},j}, v_{2,{\rm proj},j}$ & Eq.~\eqref{eq:vICRS} & Projected stellar velocities (ICRS) \\
$r_{{\rm proj},j}$ & Eq.~\eqref{eq:sproj} & Model projected separation \\
$\Delta v_{r,{\rm model},j}$ & Eq.~\eqref{eq:deltavr} & Model differential RV \\
$\Delta\mu_{\alpha,{\rm model},j}$ & Eq.~\eqref{eq:dpmra} & Model differential proper motion (RA) \\
$\Delta\mu_{\delta,{\rm model},j}$ & Eq.~\eqref{eq:dpmdec} & Model differential proper motion (Dec) \\
\hline
\multicolumn{3}{c}{\textit{Likelihoods}} \\
\hline
$s_{{\rm obs},j}$ & $\mathcal{N}(s_{{\rm proj},j}, \sigma_{s,j}^2)$ & Projected separation \\
$\Delta v_{r,{\rm obs},j}$ & $\mathcal{N}(\Delta v_{r,{\rm model},j}, \sigma_{\Delta v_r,j}^2)$ & Differential radial velocity \\
$\Delta\mu_{\alpha,{\rm obs},j}$ & $\mathcal{N}(\Delta\mu_{\alpha,{\rm model},j}, \sigma_{\mu,j}^2)$ & Differential proper motion in Right Ascension \\
$\Delta\mu_{\delta,{\rm obs},j}$ & $\mathcal{N}(\Delta\mu_{\delta,{\rm model},j}, \sigma_{\mu,j}^2)$ & Differential proper motion in Declination \\
\hline
\end{tabular}}
\end{table*}

After defining the MOND orbital dynamics and geometry, we constructed a hierarchical Bayesian model to infer the MOND parameter $a_0$ and the orbital parameters for all 57 binary systems. Previous studies relying on ensemble statistics of Gaia proper motions have reached differing conclusions, possibly due to variations in sample selection and statistical methodology \citep{Hernandez2024review}. This motivated us to construct a hierarchical Bayesian framework that simultaneously models individual orbital elements and the global MOND parameter $a_0$, propagating uncertainties and accounting for degeneracies between parameters. The graphical structure of the model is shown in Figure~\ref{fig:mond_diagram}. All priors and likelihoods are summarized in Table~\ref{tab:prior}.

\subsection{Prior distribution}\label{sec:priors}

We chose weakly informative priors that reflect physical constraints and existing knowledge while allowing the data to dominate the posterior. Below we describe and justify each prior choice:

\begin{itemize}
    \item $a_0$ : A global MOND parameter $a_0$ that applies for all 57 wide-binary pairs. We place a uniform prior on $\log_{10}a_0$ spanning from $-12$ to $-8$. We chose this range so that the canonical MOND value of $\log_{10} a_0 \approx -9.92$ sits in the mid-range of the uniform prior and we can allow for similar weights on both Newtonian and MOND regime. Going above $-8$ will not be physical as we do not expect to have such deep MOND regime. We also cannot have $a_0 < 0$ or $a_0 \approx 0$ in the prior because the MOND dynamics depend on the ratio $x = a_{\rm tot}/a_0$ when evaluating the interpolating function $\mu(x) = x / (1+x^b)^{(1/b)}$. If $a_0 \approx 0$, this ratio diverges, and the MOND acceleration becomes undefined. If $a_0 < 0$, the argument of $\mu(x)$ becomes negative which produces non-physical accelerations and complex-valued velocities in Eqs.~(\ref{eq:aeff}--\ref{eq:v2orb}). Such values cause the likelihood function to break numerically and violate the physical assumptions of MOND. We also verify the robustness of our results to alternative prior choices in Appendix~\ref{appendix:prior_sensitivity}.

    \item $M$ : Mass for each star in the binary pairs. We use Gaia FLAME masses for all 57 pairs. We adopt Gaussian priors centered on the FLAME value with the FLAME uncertainty.

    \item $a$ : Semi-major axis of the relative orbit. We parameterize  {$a = r_\perp \exp(\eta)$}, where $r_\perp$ is the observed projected separation and $\eta = \ln(a/r_\perp)$ is a free parameter for each system{, and adopt the broken power-law prior on $a$ of \mbox{\cite{Pittordis2018}}\hskip0pt,
    \begin{equation}
    P(a)\,\mathrm{d}a \propto \begin{cases} a^{-1}\,\mathrm{d}a & a \leq 5\,\mathrm{kAU} \\ a^{-1.6}\,\mathrm{d}a & a > 5\,\mathrm{kAU} \end{cases}
    \label{eq:bpl_prior}
    \end{equation}} 
     {which follows \"{O}pik's law below the break and the steeper slope measured for wide companions above it \mbox{\citep{Andrews2017, Pittordis2018}}\hskip0pt. We sample $\eta \sim \mathcal{U}(-1, 2.5)$, which is flat in $\ln a$ and therefore equivalent to \"{O}pik's law, and add $-0.6\,\ln(a/5\,\mathrm{kAU})$ to the log-prior of each system with $a > 5\,\mathrm{kAU}$ to obtain Eq.~\eqref{eq:bpl_prior}, with continuity at the break}.

    \item $e$ : Eccentricity of the relative orbit. Eccentricity distribution is important for wide-binary dynamics because orbital velocities depend strongly on eccentricity and orbital phase. High-eccentricity orbits spend most of their time near apastron with low velocities, so the observed velocity distribution is sensitive to the underlying eccentricity population. We therefore adopt a hierarchical, separation-dependent eccentricity distribution following \cite{Hwang2022eccentricity}. Binaries are divided into five separation bins based on their projected separations: $[0, 100)$, $[100, 300)$, $[300, 1000)$, $[1000, 3000)$, and $[3000, 10^6)$ AU.

    For each bin $k$, we define a population-level shape parameter:
    \begin{equation}
    \alpha_k \sim \mathcal{N}(\mu_{\alpha,k}, \sigma_{\alpha,k}^2)
    \label{eq:prior_alpha}
    \end{equation}
    where $\mu_\alpha = [-0.2, 0.4, 0.8, 1.0, 1.2]$ and $\sigma_\alpha = [0.3, 0.2, 0.2, 0.2, 0.2]$ for the five bins from the \cite{Hwang2022eccentricity}. These hyperparameters reflect the expectation that wider-binaries have higher eccentricities on average. For a system $j$ in bin $k$, the eccentricity is drawn from,
    \begin{equation}
    e_j \sim \mathrm{Beta}(\alpha_k',\, 1)
    \label{eq:prior_e}
    \end{equation}
    and truncated at $e_j < 0.98$ to avoid numerical instabilities near parabolic orbits. For this, we note that most wide binaries have extremely eccentric orbit \citep{Hwang2022extreme-eccentric}. This prior is chosen based on observational evidences \citep{Hwang2022eccentricity}. We verified that our results are not sensitive to this choice by re-running the analysis with a uniform eccentricity prior: the posterior on $a_0$ shifted by less than 0.2~dex.

    \item $i$ : Inclination of the relative orbit. We defined a uniform prior on $\cos i$ to produce an isotropic distribution for inclinations.

    \item $\omega$ : Argument of pericenter of the relative orbit. We defined $\omega$ to be uniform between $0$ and $2\pi$.

    \item $\Omega$ : Longitude of ascending node of the relative orbit. We defined $\Omega$ to be uniform between $0$ and $2\pi$.

    \item $M$ : Mean anomaly of the relative orbit. We defined $M$ to be uniform between $0$ and $2\pi$.

 \end{itemize}

\subsection{Likelihood functions}\label{sec:likelihood}
We combine constraints from the observed projected separations, differential RV, and the differential proper motion.

\begin{itemize}
    \item Projected Separation:  The observed projected separation for system $j$ is modeled as,
    \begin{equation}
    s_{\rm obs,j} \sim \mathcal{N}(s_{\rm proj,j}, \sigma_{s,j}^2)
    \label{eq:lik_sproj}
    \end{equation}
    where $\sigma_{s,j}$ is derived from \textit{Gaia} parallax uncertainties with a minimum floor of 5\% of the observed separation to account for potential systematic uncertainties in the projection geometry and unmodeled astrometric correlations.

    \item Differential Radial Velocity:  We model the differential RVs as following:
    \begin{equation}
    \Delta v_{r,\rm obs,j} \sim \mathcal{N}(\Delta v_{r,\rm model,j}, \sigma_{\Delta v_r,j}^2)
    \label{eq:lik_drv}
    \end{equation}
    where $\sigma_{\Delta v_r,j}$ is the measurement uncertainty from our differential RV analysis, obtained through bootstrap resampling of the per-order median as described in Section~\ref{sec: Data}.
    
    \item Differential Proper Motion: The proper motion components are modeled as,
    \begin{subequations}
    \begin{align}
    \Delta\mu_{\alpha,\rm obs,j} &\sim \mathcal{N}(\Delta\mu_{\alpha,\rm model,j}, \sigma_{\mu,j}^2) \label{eq:lik_pmra}\\
    \Delta\mu_{\delta,\rm obs,j} &\sim \mathcal{N}(\Delta\mu_{\delta,\rm model,j}, \sigma_{\mu,j}^2) \label{eq:lik_pmdec}
    \end{align}
    \end{subequations}
    where $\sigma_{\mu,j}$ is the \textit{Gaia} proper motion uncertainty.
\end{itemize}

\subsection{Inference with HMC}

We performed Bayesian inference using Hamiltonian Monte Carlo (HMC) using the No-U-Turn Sampler (NUTS) \citep{hoffman2014no}. We implemented this using PyMC \citep{pymc}. HMC uses gradient information to explore high-dimensional parameter spaces, making it well-suited for our model with thousands of parameters including orbital elements for all 57 binary systems and population-level hyperparameters.

The NUTS algorithm automatically tunes the step size and number of leapfrog steps during an initial tuning phase by adapting to the local geometry of the posterior distribution. We ran 4 independent Markov chains, each with 2000 tuning steps and 3000 sampling steps, which produced a total of 12000 posterior samples. We set the target acceptance probability to 0.95 and the maximum tree depth to 10. This acceptance probability is standard for NUTS when dealing with correlated posteriors. Unlike traditional Metropolis-Hastings samplers where high acceptance rates can indicate poor mixing, NUTS automatically adapts its trajectory length based on the local posterior geometry. The higher target acceptance probability reduces the step size, which helps navigate the strong degeneracies between orbital parameters like semi-major axis $a$, eccentricity $e$, and mean anomaly $M$ without overshooting.

\subsection{Model Diagnostics}\label{sec:diagnostics}

We used multiple diagnostics to assess convergence. The primary method was the Gelman-Rubin statistic $\hat{R}$ \citep{gelman1992inference}, which compares the between-chain and within-chain variances to test convergence of a model. For our study,  we required $\hat{R} < 1.01$ for all parameters to ensure that the chains have converged to the same distribution. In our final analysis, all parameters satisfied this requirement.

We also examined the distribution of residuals to identify potential model misspecifications or outliers. For the differential RV likelihood, we computed normalized residuals,
\begin{equation}
z_i = \frac{\Delta v_{r,\rm obs,i} - \Delta v_{r,\rm model,i}}{\sigma_{\Delta v_r,i}}
\end{equation}

Under a well-specified model, these residuals should follow a standard normal distribution. We found that the residual distribution has mean $\approx 0$ and standard deviation $\approx 1.1$, corresponding to a reduced chi-square of $\chi^2_\nu \approx 1.2$. This indicates a good model fit with slight overdispersion that may reflect unmodeled systematics or intrinsic stellar variability.

\begin{figure*}
    \centering
        \begin{overpic}[width=1\textwidth]{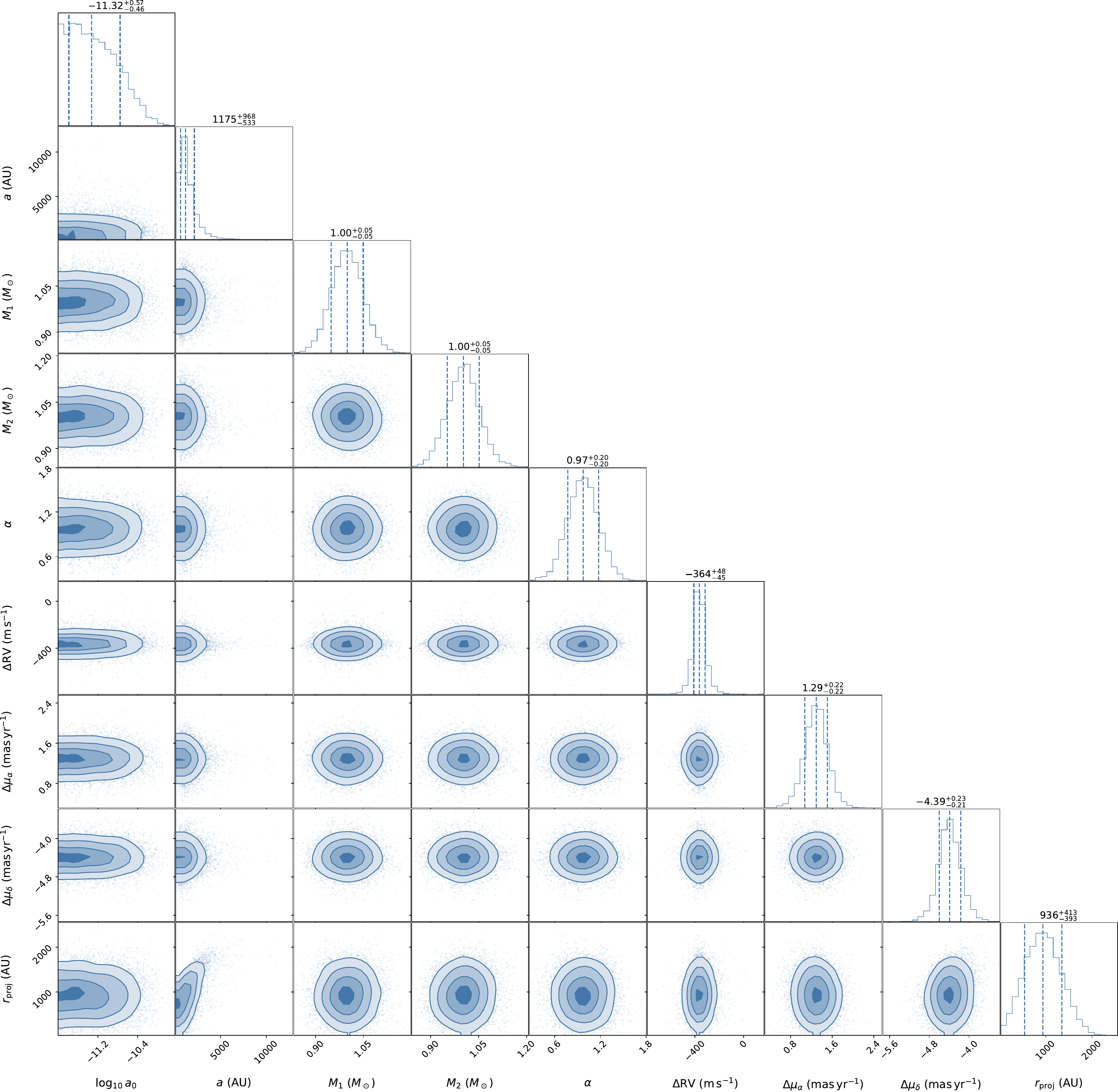}
        \end{overpic}
        \caption{Corner plot showing the posterior distributions for one representative binary system, Gaia 6644929656684912512 and Gaia 6644929416166727680 considering $b=1$. The parameters shown are: MOND acceleration scale in log-scale $\log_{10} a_0$, semi-major axis $a$, stellar masses $M_1$ and $M_2$, eccentricity distribution power $\alpha$, model differential RV $\Delta$RV, model differential proper motions $\Delta\mu_\alpha$ and $\Delta\mu_\delta$, and model projected separation $r_{\rm proj}$. We note that, though this figure shows the posterior for a single system, the inference was performed jointly across all 57 systems, with $\log_{10}a_0$ shared globally. The correlations shown here reflect the degeneracies for this particular system marginalized over the full joint posterior.}
        \label{fig:corner94}
\end{figure*}

Figure~\ref{fig:corner94} shows a corner plot of the posterior for one representative binary pair for $b=1$, showing the correlations between $a_0$, orbital elements, and observables. The posteriors show good convergence visually with no apparent trends or multi-modality. Additional corner plots for other systems and for $b=2$ are provided in Appendix~\ref{appendix:corner}.

\subsection{Posterior Analysis}

\begin{figure}
    \centering
        \begin{overpic}[width=0.48\textwidth]{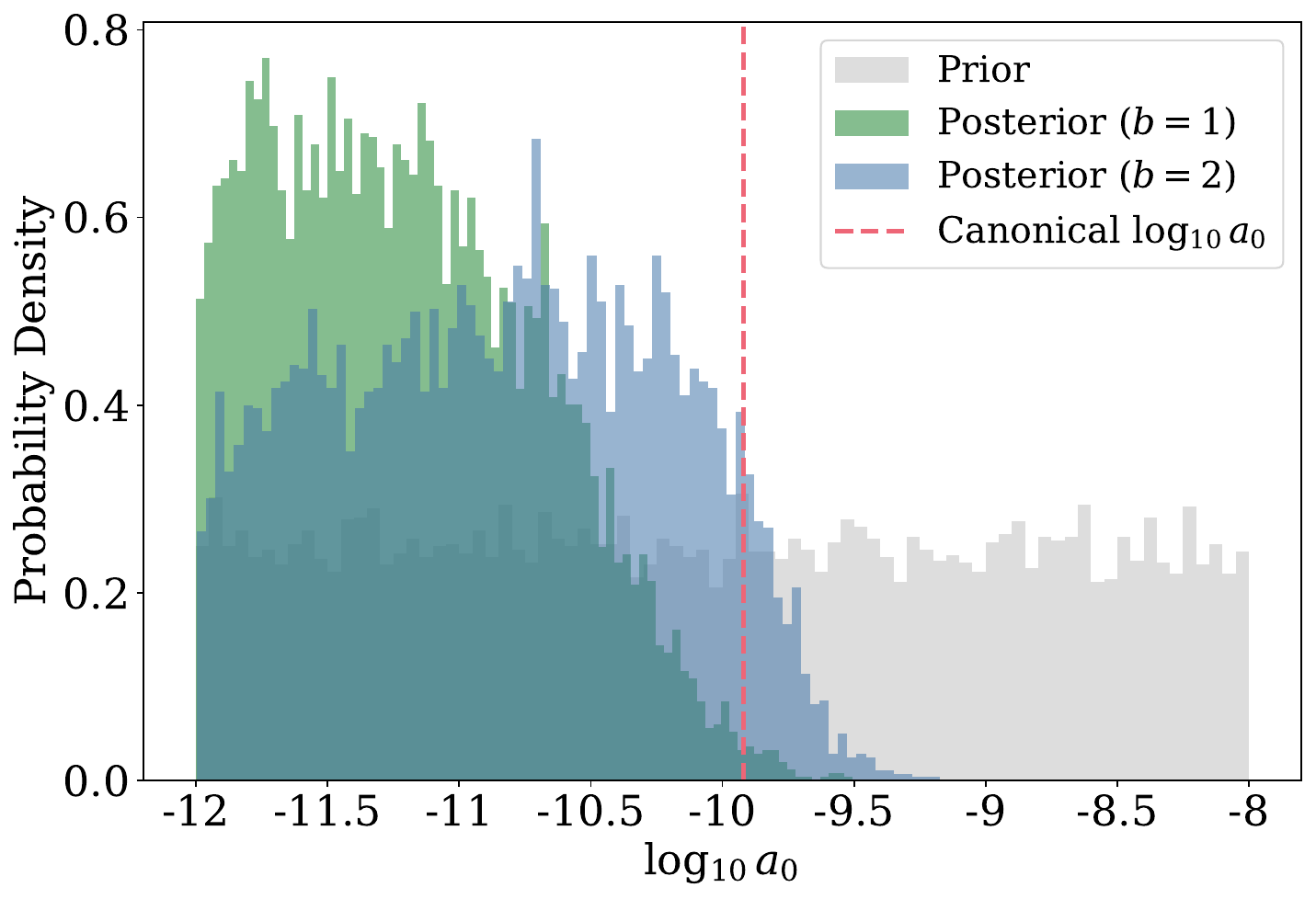}
        \end{overpic}
        \caption{Comparison of posterior distributions for $\log_{10} a_0$ using the simple ($b=1$, green) and standard ($b=2$, blue) interpolating functions. The gray histogram shows the uniform prior $\mathcal{U}(-12, -8)$. The vertical dashed red line marks the canonical MOND value $\log_{10} a_0 = -9.92$. The $b=2$ posterior is shifted toward higher $a_0$ values, indicating that constraints on $a_0$ depend on the assumed form of the interpolating function.}
        \label{fig:posterior_b12}
\end{figure}

\begin{table}
\caption{Posterior constraints on $\log_{10} a_0$ for the two interpolating functions using the uniform prior $\mathcal{U}(-12, -8)$. These results are derived from the 57 selected bound systems, all of which reside in the MOND transition regime. Results for alternative prior choices are presented in Appendix~\ref{appendix:prior_sensitivity}.}
\label{tab:results}
\centering
\begin{tabular}{lccc}
\hline
 & & Posterior & Equivalent  \\
 & & CDF at & Gaussian  \\
$b$ & Median & Canonical & Exclusion \\
\hline
1 & $-11.24$ & 99.4\% & $2.5\sigma$ \\
2 & $-10.83$ & 92.9\% & $1.5\sigma$ \\
\hline
\end{tabular}
\end{table}

The primary output of our analysis is the posterior distribution on $\log_{10} a_0$ for different choices of interpolating functions. We ran our hierarchical model with two fixed values of the sharpness parameter: $b=1$ (simple interpolating function) and $b=2$ (standard interpolating function). Figure~\ref{fig:posterior_b12} shows the posterior distributions for both interpolating functions using our fiducial uniform prior $\mathcal{U}(-12, -8)$.

Table~\ref{tab:results} summarizes our constraints on $a_0$ for both interpolating functions. For $b=1$, the posterior has a median of $\log_{10} a_0 = -11.24$ with a 68\% credible interval of $[-11.75, -10.65]$. The canonical MOND value of $\log_{10} a_0 = -9.92$ lies at the 99.4th percentile of the posterior distribution, corresponding to exclusion at $2.5\sigma$. Given our adopted treatment of the external field effect and the $b=1$ interpolating function, this represents tension with MOND at the accepted $a_0$ value.

For $b=2$, the posterior is broader and shifted toward higher values of $a_0$, with a median of $\log_{10} a_0 = -10.83$ and a 68\% credible interval of $[-11.58, -10.15]$. The canonical MOND value lies at the 92.9th percentile, corresponding to exclusion at $1.5\sigma$. This weaker tension for $b=2$ reflects the fact that the standard interpolating function produces a more gradual transition between regimes.

Because all of our systems reside in the transition regime, the constraints on $a_0$ are sensitive to the choice of interpolating function. This is expected: in the transition regime, the interpolating function determines the mapping between acceleration and velocity boost. Different interpolating functions therefore lead to different inferred values of $a_0$ for the same observed kinematics.

Figure~\ref{fig:posterior_b12} directly compares the two posteriors. The $b=2$ posterior is systematically shifted toward higher $a_0$ values compared to $b=1$. It shows that constraints on $a_0$ from wide-binary dynamics depend on the assumed form of the interpolating function.

We verify that these results are robust to the choice of prior in Appendix~\ref{appendix:prior_sensitivity}, where we present results for two alternative prior ranges: $\mathcal{U}(-11, -9)$ (narrower) and $\mathcal{U}(-13, -7)$ (wider). The qualitative conclusions remain unchanged across all prior choices.

As a supplementary test, we also inferred an effective gravity boost factor $\gamma$ for the  C3PO systems  {that are bound in three dimensions, split at a projected separation of 3000~AU, finding $\gamma = 1.17^{+0.37}_{-0.23}$ for the 18 systems above and $1.05^{+0.11}_{-0.08}$ for the 39 below, both} consistent with Newtonian gravity ($\gamma = 1$). However, this measurement is less informative than the direct $a_0$ inference for our sample, because the C3PO systems reside in the MOND transition regime where the MOND predicted boost varies (see Appendix~\ref{appendix:gamma_test} for the full analysis).

We also confirm that the posterior distribution of $\cos i$ is consistent with isotropy and that the mean anomaly distribution is consistent with uniformity, as expected for gravitationally bound systems observed at random times (Appendix~\ref{appendix:prior_sensitivity}).

\section{Discussion} \label{sec: Discussion}
\subsection{Results and statistical power}

Our differential RV technique achieves per-order precisions of $20$--$40~\rm m\,s^{-1}$, which when combined across $\sim$30--50 echelle orders yields final precisions of $\sim$8--15~m\,s$^{-1}$ per binary pair. This is comparable to the stability of exoplanet RV spectrographs such as HARPS ($\sim1~\rm m\,s^{-1}$) and HIRES with iodine cells ($\sim3~\rm m\,s^{-1}$), but achieved without specialized calibration hardware. Our technique is applicable to stellar pairs with similar spectral types, where one star serves as a spectral template for the other. This enables echelle-order-by-order comparison that cancels instrumental systematics.

Our sample of 57 binaries provides stronger constraints than studies using tens of thousands of \textit{Gaia} wide binaries, which reflects the trade-off between sample size and measurement precision. Since Fisher information scales as $\sigma^{-2}$, our $\sim$24-fold precision improvement outweighs the smaller sample size. Beyond this, the measurement precision also determines which statistical approaches are viable. For \textit{Gaia} RVs, the expected differential velocities of 50--200~m\,s$^{-1}$ are smaller than typical uncertainties, meaning each system has signal-to-noise $\ll 1$. In this regime, individual orbital parameters are poorly constrained, so \textit{Gaia}-based studies rely on ensemble statistics, comparing the aggregate distribution of observed velocities to forward-simulated predictions under different gravity models \citep{El-badry2019}. Our differential RV precision yields signal-to-noise $\approx$ 2--7 per system, which enables us to constrain individual orbits and perform hierarchical Bayesian inference that jointly fits orbital elements and the global parameter $a_0$. This approach can identify outliers and propagate orbital degeneracies into the final uncertainty on $a_0$.

Furthermore, as Fisher information scales as $\sigma^{-2}$, our 57 systems carry effective statistical weight equivalent to several $\times10^4$ Gaia-quality systems. This is more than the total number of F, G, and K-type wide binaries available in \textit{Gaia} at comparable separations ($\sim$1,000--10,000 AU). Both of these reasons explain why our smaller sample has tighter constraints on $a_0$ than \textit{Gaia}.

Our analysis yields posterior distributions for $\log_{10} a_0$ that depend on the assumed form of the MOND interpolating function. For $b=1$, the canonical MOND value $a_0 = 1.2 \times 10^{-10}$ m\,s$^{-2}$ lies at the 99.4th percentile of the posterior, excluded at $2.5\sigma$. For $b=2$, the canonical value lies at the 92.9th percentile, excluded at $1.5\sigma$.

\subsection{Interpretations}

Several interpretations of our results are possible which are described in the following:

\begin{itemize}
    \item Most of the wide-binary systems we used for this study reside in the MOND transition regime ($0.1 < a_N/a_0 < 10$). This means our test probes the regime where the interpolating function plays a significant role and where the treatment of the external field effect is most uncertain. The sensitivity to interpolating function choice, with $b=1$ showing $2.5\sigma$ tension and $b=2$ showing $1.5\sigma$ tension, directly reflects this.

    \item Our analytic treatment of the external field effect (Eqs.~(\ref{eq:atot}--\ref{eq:aeff})) is an approximation. \cite{PflammAltenburg2025} has shown through exact numerical integration that these approximate expressions overestimate the MOND velocity boost factor in the transition regime. As all of our systems reside in this regime, this systematic effect biases our inferred $a_0$ toward lower values. The direction of the bias is toward lower $a_0$: correcting for it would shift the inferred $a_0$ upward toward the canonical value, so our reported exclusion levels ($2.5\sigma$ for $b=1$ and $1.5\sigma$ for $b=2$) are upper bounds on the true tension with MOND. Correcting for this would require either adopting the numerical results from \cite{PflammAltenburg2025} or implementing exact numerical integration.

    \item Our results could indicate that MOND, at least in its standard formulation, does not accurately predict the gravitational interaction between wide-binaries embedded in an external galactic field. This would not necessarily invalidate MOND's success on galactic scales if MOND is viewed as an effective description rather than fundamental physics, but it would represent a limitation of the theory. 
    
    \item The value $a_0 = 1.2 \times 10^{-10}$ m\,s$^{-2}$ was determined primarily from rotation curves of spiral galaxies spanning 10--100 kpc with masses of $10^{10}$--$10^{12}$ $\rm M_\odot$ \citep{Begeman1991, McGaugh2011MOND, Famaey2012}. It is possible that $a_0$ depends on the physical environment, mass scale, or system type in ways not captured by the standard MOND framework. If the true uncertainty on $a_0$ is larger than commonly assumed, our measurements would be less discrepant than they appear. We note that our inference is conducted in $\log_{10}(a_0)$ space, which necessarily excludes $a_0 = 0$. This is appropriate for testing MOND, where $a_0 > 0$ by construction; a direct comparison with Newtonian gravity would require a separate model comparison framework. The tendency of the posterior to accumulate toward lower $a_0$ values is suggestive, and we have verified this trend is robust across different prior choices (Appendix~\ref{appendix:prior_sensitivity}).
    
    \item Our results show sensitivity to the choice of interpolating function. The $b=2$ posterior shows weaker tension with the canonical $a_0$ than $b=1$. It is possible that the true interpolating function differs from both standard forms, or that additional parameters beyond $a_0$ and $b$ are needed to describe MOND dynamics in external fields. We do not have enough data, but if we had, it would be possible to jointly fit $b$ and $a_0$ to determine which interpolating functions are preferred by the observations. 

    \item Our treatment of the external field effect assumes simple vector addition of internal and external accelerations, which may not fully capture the behavior predicted by different MOND formulations. The precise behavior of MOND in external fields remains an area of active development, and relativistic extensions such as TeVeS may predict different behaviors \citep{Bekenstein2004MONDext, Blanchet2011MONDext, Sanders2015MONDext, Chae2022ext}. We also assumed a constant external field pointing toward the Galactic center with magnitude $a_{\rm ext} = 2.1 \times 10^{-10}$ m\,s$^{-2}$, but the true external field may include system-to-system variations due to local density fluctuations.

\end{itemize}

\subsection{Comparison to previous studies}

Our approach differs from previous wide-binary MOND tests in several ways. Most importantly, we directly infer the MOND acceleration scale $a_0$ as a free parameter, rather than assuming its canonical value and testing for consistency. Previous wide-binary studies using \textit{Gaia} data \citep{Hernandez2022, Hernandez2023, Hernandez2024, Chae2023, Chae2024a, Chae2024b, Pittordis2023, Banik2024} have measured the effective gravity boost $\gamma$ (where $G \rightarrow \gamma G$). The inference of $\gamma$ in these papers is a pure kinematic measurement that assumes only Keplerian orbits; it does not assume any particular value of $a_0$ or any MOND framework. These studies reported $\gamma = 1.43 \pm 0.06$ \citep{Chae2023}, $\gamma = 1.49 \pm 0.2$ \citep{Chae2024a}, $\gamma = 1.5 \pm 0.2$ \citep{Hernandez2024}, and $\gamma \approx 1.6 \pm 0.2$ \citep{Chae2026}, which can be compared to AQUAL predictions of $\gamma \approx 1.4$ for wide binaries in the solar neighborhood. In contrast, \cite{Banik2024} reported a strong preference for Newtonian gravity ($\gamma = 1$).

Our $\gamma$ test on the C3PO sample (Appendix~\ref{appendix:gamma_test}) yields  {$\gamma = 1.17^{+0.37}_{-0.23}$ for the 18 systems above 3000~AU that are bound in three dimensions and $1.05^{+0.11}_{-0.08}$ for the 39 below, both} consistent with Newtonian gravity. As discussed in Appendix~\ref{appendix:gamma_test}, because most of our systems reside in the MOND transition regime, the effective gravity boost is not expected to be constant across the sample, making a single global $\gamma$ a less natural parameterization than $a_0$ for this dataset.

In the companion paper \citep{SaadTing2026}, we applied the same framework to the \cite{Chae2026} dataset, which includes systems in the deep MOND regime, and found $\gamma = 1.12^{+0.27}_{-0.22}$, again consistent with Newtonian gravity. We also showed that the inferred value of $\gamma$ is sensitive to the orbital modeling assumptions, in particular whether the semi-major axis is treated as an independent parameter or derived from a geometric de-projection of the projected separation.

As discussed in the review by \cite{Hernandez2024review}, the conflicting conclusions in previous studies arise from multiple factors. A fundamental methodological concern with \cite{Banik2024} is that while observational uncertainties in the relative velocities were used to select the sample, these uncertainties played no role in the inference procedure itself, and nor in the determination of confidence intervals on the final results. This led to the 19$\sigma$ claims that have been questioned by subsequent analyses. In contrast, the analyses by Hernandez, Chae, and collaborators incorporate velocity uncertainties throughout their statistical procedures.

Our approach is more analogous to how $a_0$ was originally determined from galaxy rotation curves \citep{Begeman1991, McGaugh2011MOND}: we treat $a_0$ as a free parameter and ask what value best explains the observations. This is a direct test of whether the same $a_0$ that governs galactic dynamics also applies to wide binary systems. The fact that our inferred $a_0$ values lie below the canonical value suggests possible tension between wide-binary kinematics and the MOND acceleration scale derived from galaxies, though we caution that this conclusion depends on the accuracy of our external field effect treatment.

Beyond this difference, our approach uses high-precision differential RVs and hierarchical Bayesian inference to model each binary system in three dimensions. Previous \textit{Gaia}-based studies relied primarily on sky-projected (2D) relative velocities derived from proper motions, as \textit{Gaia} RV uncertainties ($\sim$1000~m\,s$^{-1}$) exceed the expected velocity differences in wide binaries. The most directly comparable work is \cite{Saglia2025}, who used archival HARPS observations of 32 wide binaries to perform three-dimensional kinematic analysis with precise RVs. However, their study tested only Newtonian gravity without fitting for MOND parameters, measured absolute rather than differential RVs (increasing sensitivity to instrumental systematics and stellar convective shifts), and used a smaller sample. We note that our model framework is general and can be applied to systems in any acceleration regime. Systems in the Newtonian regime do not bias the model toward Newtonian gravity; rather, they contribute less information to the global inference of $a_0$ because MOND and Newtonian predictions converge at high accelerations. The constraining power comes primarily from systems in or near the transition regime where the theories differ. We also note that it may be possible to use systems in the transition regime to test different external field effect formulations by fixing $a_0$ at its canonical value and asking which treatment best reproduces the observed kinematics. This would require modifications to our model framework and is beyond the scope of the current work.

\subsection{Methodological considerations}
\subsubsection{Data and measurements}

An improvement in getting more precise RV is to use full Bayesian forward modeling of each spectrum including instrumental line-spread functions, telluric contamination, and order-level jitter terms \citep{Langrford2024rv}. While this could yield slightly tighter uncertainties, performing such fits for 200 stars individually is computationally expensive. Another direction of improvement is the easiest to guess: increasing the sample size by collecting more high resolution and high SNR data of wide-binaries. To verify that our conclusions are not driven by a few systems with unusually high precision, we repeated our analysis excluding high-precision outliers and find consistent results (our robustness analysis).

As discussed in Section~\ref{sec: Data}, the C3PO selection criteria ensure that both components of each pair have similar stellar parameters, which suppresses astrophysical differential RV offsets from convective blueshift and gravitational redshift to levels well below our measurement uncertainties. The slight overdispersion in the residuals ($\chi^2_\nu \approx 1.2$) may partly reflect these effects, but their contribution is negligible at the order-of-magnitude level.

A potential concern is that stars with different surface gravities or effective temperatures can exhibit systematic differential RV offsets due to convective blueshift and gravitational redshift. The gravitational redshift for a solar-type star is $v_{\rm grav} \approx 636(M/{\rm M}_\odot)(R_\odot/R)$ m\,s$^{-1}$ \citep{Gray2009}, so the differential contribution scales with the difference in $M/R$ between the two stars. For our sample, the C3PO selection criteria ($|\Delta(\mathrm{BP-RP})| < 0.15$ mag) ensure that both components have similar stellar parameters. Using \textit{Gaia} FLAME masses and radii, the range of $\Delta(M/R)$ across our sample is $\lesssim 0.5$, corresponding to a differential gravitational redshift of order $\sim1$ m\,s$^{-1}$. Similarly, convective blueshift varies with spectral type \citep{Meunier2017}, but the color similarity requirement restricts the range over which this effect changes appreciably; we estimate the differential convective blueshift contribution to be $\lesssim0.1$ m\,s$^{-1}$ across the sample. Both contributions are an order of magnitude smaller than our typical differential RV uncertainties of $\sim8$--15 m\,s$^{-1}$ and are therefore negligible for our analysis.

\subsubsection{Modeling}\label{sec:disc:modeling}

The systemic velocities of the binary systems are currently modeled with independent Gaussian priors centered at zero for each of the three velocity components. These systemic velocities represent the center-of-mass motion of each binary through the Galaxy, which must be separated from the orbital velocities we use to constrain $a_0$. Our current approach is appropriate for the thin disk population that dominates our sample, but it does not exploit the fact that nearby stars exhibit kinematic structure related to the local standard of rest and known moving groups. Incorporating a kinematic model that accounts for the velocity distribution of the thin disk could help constrain the systemic velocities more tightly, reducing degeneracies with the orbital velocities.

Our prior on $\log_{10}a_0$ is uniform from $-12$ to $-8$. This range was chosen so that the canonical MOND value ($\log_{10}a_0=-9.92$) lies near the middle to allow the data to favor either Newtonian-like behavior (lower $a_0$) or MOND-like behavior (higher $a_0$) with similar prior weights. As discussed in our formalism, the model requires a non-zero $a_0$. However, this also means that our conclusions depend on this prior choice, especially if the true value of $a_0$ is zero (and hence $\log_{10} a_0$ approaches negative infinity). We deem our choice of prior reasonable given the current formalism, but ultimately, the constraints suggest that the data would benefit from a larger sample size. While there are tentative suggestions that the posterior is concentrating toward the lower end of the allowed prior range, a larger sample would mitigate the current dependency on prior choice, and expanding the C3PO sample would be beneficial for this purpose. We present results for alternative prior choices in Appendix~\ref{appendix:prior_sensitivity}.

\section{Conclusions} \label{sec: Conclusions}

We present a test of MOND using high-precision differential RVs of 57 gravitationally bound wide-binary pairs from the C3PO survey. Firstly, we developed a pixel-integrated spectral fitting technique for echelle spectroscopy that achieved differential RV precisions of $\sim$8--15~m\,s$^{-1}$ per binary pair, a factor of $\sim$24 improvement over \textit{Gaia} RVs.

Secondly, we applied a bound-orbit selection criterion based on the scaled velocity $\tilde{v} < 2.5$, following \cite{Cookson2026}, to remove likely chance alignments from our initial sample of 85 pairs with differential RV measurements. This yielded 57 gravitationally bound systems. We determined that all 57 systems reside in the MOND transition regime ($0.1 < a_N/a_0 < 10$), with none in the deep MOND regime. Our test is therefore performed entirely in the transition regime, where the choice of interpolating function plays a significant role and where the approximate treatment of the external field effect is least accurate.

Thirdly, we constructed a hierarchical Bayesian model that jointly fits orbital elements for all 57 binary systems and the global MOND acceleration scale $a_0$, using MOND-modified Keplerian dynamics that include the external galactic field.

Our results suggest tension with MOND at the presently accepted $a_0$ value, given our adopted treatment of the external field effect and interpolating functions. For the simple interpolating function ($b=1$), the posterior median is $\log_{10}a_0 = -11.24$, with the canonical MOND value ($\log_{10}a_0 = -9.92$) excluded at $2.5\sigma$. For the standard interpolating function ($b=2$), the exclusion is weaker at $1.5\sigma$, with a posterior median of $\log_{10}a_0 = -10.83$. The dependence of our results on the interpolating function is expected given that all systems are in the transition regime.

Because the separations and masses of the C3PO binaries imply accelerations near the MOND transition regime ($a_N \sim a_0$), our constraints are sensitive to the assumed interpolating function and the treatment of the external field effect. Within the MOND framework, our findings suggest either that the canonical $a_0$ value does not apply universally across astrophysical systems, that the interpolating function differs from standard forms, or that the external field effect in the solar neighborhood requires a more accurate treatment. The methodology developed here can be applied to larger samples of wide binaries to provide increasingly precise tests of gravity in the low-acceleration regime.

Looking forward, expanding the sample to include deep MOND systems ($a_N/a_0 < 0.1$) would provide a more definitive test of MOND, as the gravitational anomaly is more pronounced and less sensitive to the interpolating function in this regime. Such systems require observations of binaries at larger separations ($\gtrsim$10 kAU) with similar spectral precision, which is achievable with dedicated observing programs on 4--10m class telescopes.

\section*{Acknowledgements}
We thank Kareem El-Badry, Indranil Banik, Xavier Hernandez, and Kyu-Hyun Chae for helpful discussions. We
also appreciate the comments on precise RV measurements from Andrew Gould, Alex Ji, and Daniel Kelson
throughout the past few months. We thank Fan Liu for
helping curate the data as part of the C3PO program.
We are grateful to the two anonymous referees for their constructive reports.

SMS is supported by the Distinguished University Fellowship awarded by The Ohio State University. YST is
supported by NSF under Grant AST-2406729.

This work has made use of data from the European Space Agency (ESA) mission {\textit{Gaia}} (\url{https://www.cosmos.esa.int/gaia}), processed by the {\textit{Gaia}} Data Processing and Analysis Consortium (DPAC, \url{https://www.cosmos.esa.int/web/gaia/dpac/consortium}). Funding for the DPAC has been provided by national institutions, in particular the institutions participating in the \textit{Gaia} Multilateral Agreement.

While the scientific content, draft and analysis are entirely from the authors, this project has benefited from collaboration with AI tools: Cursor for coding development and Claude-Opus-4.5 for proofreading of the final manuscript.

This work has also made use of the following software packages:
AstroPy \citep{astropy},
PyMC \citep{pymc},
and ArviZ \citep{arviz}.

\section*{Data Availability}
The data and code used in this work are publicly available. The differential radial velocity measurements, the \textit{Gaia} astrometric data for all 85 C3PO wide-binary pairs, and the full analysis code including the hierarchical Bayesian model implementation can be found on GitHub at \url{https://github.com/seratsaad/wb3d-mond}.

\bibliographystyle{mnras}
\bibliography{references.bib}

\begin{thebibliography}{}
\makeatletter
\relax
\def\mn@urlcharsother{\let\do\@makeother \do\$\do\&\do\#\do\^\do\_\do\%\do\~}
\def\mn@doi{\begingroup\mn@urlcharsother \@ifnextchar [ {\mn@doi@}
  {\mn@doi@[]}}
\def\mn@doi@[#1]#2{\def\@tempa{#1}\ifx\@tempa\@empty \href
  {http://dx.doi.org/#2} {doi:#2}\else \href {http://dx.doi.org/#2} {#1}\fi
  \endgroup}
\def\mn@eprint#1#2{\mn@eprint@#1:#2::\@nil}
\def\mn@eprint@arXiv#1{\href {http://arxiv.org/abs/#1} {{\tt arXiv:#1}}}
\def\mn@eprint@dblp#1{\href {http://dblp.uni-trier.de/rec/bibtex/#1.xml}
  {dblp:#1}}
\def\mn@eprint@#1:#2:#3:#4\@nil{\def\@tempa {#1}\def\@tempb {#2}\def\@tempc
  {#3}\ifx \@tempc \@empty \let \@tempc \@tempb \let \@tempb \@tempa \fi \ifx
  \@tempb \@empty \def\@tempb {arXiv}\fi \@ifundefined
  {mn@eprint@\@tempb}{\@tempb:\@tempc}{\expandafter \expandafter \csname
  mn@eprint@\@tempb\endcsname \expandafter{\@tempc}}}

\bibitem[\protect\citeauthoryear{{Andrews}, {Chanam{\'e}}  \&
  {Ag{\"u}eros}}{{Andrews} et~al.}{2017}]{Andrews2017}
{Andrews} J.~J.,  {Chanam{\'e}} J.,   {Ag{\"u}eros} M.~A.,  2017, \mn@doi
  [\mnras] {10.1093/mnras/stx2000}, \href
  {https://ui.adsabs.harvard.edu/abs/2017MNRAS.472..675A} {472, 675}

\bibitem[\protect\citeauthoryear{{Angus}}{{Angus}}{2008}]{Angus2008}
{Angus} G.~W.,  2008, \mn@doi [\mnras] {10.1111/j.1365-2966.2008.13351.x},
  \href {https://ui.adsabs.harvard.edu/abs/2008MNRAS.387.1481A} {387, 1481}

\bibitem[\protect\citeauthoryear{{Astropy Collaboration} et~al.,}{{Astropy
  Collaboration} et~al.}{2013}]{astropy}
{Astropy Collaboration} et~al., 2013, \mn@doi [\aap]
  {10.1051/0004-6361/201322068}, \href
  {https://ui.adsabs.harvard.edu/abs/2013A&A...558A..33A} {558, A33}

\bibitem[\protect\citeauthoryear{{Banik} \& {Zhao}}{{Banik} \&
  {Zhao}}{2018}]{Banik2018}
{Banik} I.,  {Zhao} H.,  2018, \mn@doi [\mnras] {10.1093/mnras/sty2007}, \href
  {https://ui.adsabs.harvard.edu/abs/2018MNRAS.480.2660B} {480, 2660}

\bibitem[\protect\citeauthoryear{{Banik}, {Pittordis}, {Sutherland}, {Famaey},
  {Ibata}, {Mieske}  \& {Zhao}}{{Banik} et~al.}{2024}]{Banik2024}
{Banik} I.,  {Pittordis} C.,  {Sutherland} W.,  {Famaey} B.,  {Ibata} R.,
  {Mieske} S.,   {Zhao} H.,  2024, \mn@doi [\mnras] {10.1093/mnras/stad3393},
  \href {https://ui.adsabs.harvard.edu/abs/2024MNRAS.527.4573B} {527, 4573}

\bibitem[\protect\citeauthoryear{{Bedell} et~al.,}{{Bedell}
  et~al.}{2018}]{Bedell2018}
{Bedell} M.,  et~al., 2018, \mn@doi [\apj] {10.3847/1538-4357/aad908}, \href
  {https://ui.adsabs.harvard.edu/abs/2018ApJ...865...68B} {865, 68}

\bibitem[\protect\citeauthoryear{{Begeman}}{{Begeman}}{1989}]{Begeman1989}
{Begeman} K.~G.,  1989, \aap, \href
  {https://ui.adsabs.harvard.edu/abs/1989A&A...223...47B} {223, 47}

\bibitem[\protect\citeauthoryear{{Begeman}, {Broeils}  \& {Sanders}}{{Begeman}
  et~al.}{1991}]{Begeman1991}
{Begeman} K.~G.,  {Broeils} A.~H.,   {Sanders} R.~H.,  1991, \mn@doi [\mnras]
  {10.1093/mnras/249.3.523}, \href
  {https://ui.adsabs.harvard.edu/abs/1991MNRAS.249..523B} {249, 523}

\bibitem[\protect\citeauthoryear{{Bekenstein}}{{Bekenstein}}{2004a}]{Bekenstein2004}
{Bekenstein} J.~D.,  2004a, \mn@doi [\prd] {10.1103/PhysRevD.70.083509}, \href
  {https://ui.adsabs.harvard.edu/abs/2004PhRvD..70h3509B} {70, 083509}

\bibitem[\protect\citeauthoryear{{Bekenstein}}{{Bekenstein}}{2004b}]{Bekenstein2004MONDext}
{Bekenstein} J.~D.,  2004b, \mn@doi [\prd] {10.1103/PhysRevD.70.083509}, \href
  {https://ui.adsabs.harvard.edu/abs/2004PhRvD..70h3509B} {70, 083509}

\bibitem[\protect\citeauthoryear{{Bekenstein} \& {Milgrom}}{{Bekenstein} \&
  {Milgrom}}{1984}]{Bekenstein1984}
{Bekenstein} J.,  {Milgrom} M.,  1984, \mn@doi [\apj] {10.1086/162570}, \href
  {https://ui.adsabs.harvard.edu/abs/1984ApJ...286....7B} {286, 7}

\bibitem[\protect\citeauthoryear{{Bernstein}, {Shectman}, {Gunnels},
  {Mochnacki}  \& {Athey}}{{Bernstein} et~al.}{2003}]{Bernstein2003MIKE}
{Bernstein} R.,  {Shectman} S.~A.,  {Gunnels} S.~M.,  {Mochnacki} S.,   {Athey}
  A.~E.,  2003, in {Iye} M.,  {Moorwood} A. F.~M.,  eds,  Society of
  Photo-Optical Instrumentation Engineers (SPIE) Conference Series Vol. 4841,
  Instrument Design and Performance for Optical/Infrared Ground-based
  Telescopes. pp 1694--1704, \mn@doi{10.1117/12.461502}

\bibitem[\protect\citeauthoryear{{Blanchet} \& {Novak}}{{Blanchet} \&
  {Novak}}{2011}]{Blanchet2011MONDext}
{Blanchet} L.,  {Novak} J.,  2011, \mn@doi [\mnras]
  {10.1111/j.1365-2966.2010.18076.x}, \href
  {https://ui.adsabs.harvard.edu/abs/2011MNRAS.412.2530B} {412, 2530}

\bibitem[\protect\citeauthoryear{{Bouchy}, {Pepe}  \& {Queloz}}{{Bouchy}
  et~al.}{2001}]{Bouchy2001}
{Bouchy} F.,  {Pepe} F.,   {Queloz} D.,  2001, \mn@doi [\aap]
  {10.1051/0004-6361:20010730}, \href
  {https://ui.adsabs.harvard.edu/abs/2001A&A...374..733B} {374, 733}

\bibitem[\protect\citeauthoryear{{Bullock} \& {Boylan-Kolchin}}{{Bullock} \&
  {Boylan-Kolchin}}{2017}]{Bullock2017}
{Bullock} J.~S.,  {Boylan-Kolchin} M.,  2017, \mn@doi [\araa]
  {10.1146/annurev-astro-091916-055313}, \href
  {https://ui.adsabs.harvard.edu/abs/2017ARA&A..55..343B} {55, 343}

\bibitem[\protect\citeauthoryear{{Chae}}{{Chae}}{2023}]{Chae2023}
{Chae} K.-H.,  2023, \mn@doi [\apj] {10.3847/1538-4357/ace101}, \href
  {https://ui.adsabs.harvard.edu/abs/2023ApJ...952..128C} {952, 128}

\bibitem[\protect\citeauthoryear{{Chae}}{{Chae}}{2024a}]{Chae2024b}
{Chae} K.-H.,  2024a, \mn@doi [\apj] {10.3847/1538-4357/ad0ed5}, \href
  {https://ui.adsabs.harvard.edu/abs/2024ApJ...960..114C} {960, 114}

\bibitem[\protect\citeauthoryear{{Chae}}{{Chae}}{2024b}]{Chae2024a}
{Chae} K.-H.,  2024b, \mn@doi [\apj] {10.3847/1538-4357/ad61e9}, \href
  {https://ui.adsabs.harvard.edu/abs/2024ApJ...972..186C} {972, 186}

\bibitem[\protect\citeauthoryear{{Chae}, {Lelli}, {Desmond}, {McGaugh}  \&
  {Schombert}}{{Chae} et~al.}{2022}]{Chae2022ext}
{Chae} K.-H.,  {Lelli} F.,  {Desmond} H.,  {McGaugh} S.~S.,   {Schombert}
  J.~M.,  2022, \mn@doi [\prd] {10.1103/PhysRevD.106.103025}, \href
  {https://ui.adsabs.harvard.edu/abs/2022PhRvD.106j3025C} {106, 103025}

\bibitem[\protect\citeauthoryear{{Chae}, {Lee}, {Hernandez}, {Orlov}, {Lim},
  {Turnshek}  \& {Lee}}{{Chae} et~al.}{2026}]{Chae2026}
{Chae} K.-H.,  {Lee} B.-C.,  {Hernandez} X.,  {Orlov} V.~G.,  {Lim} D.,
  {Turnshek} D.~A.,   {Lee} Y.-W.,  2026, \mn@doi [arXiv e-prints]
  {10.48550/arXiv.2601.21728}, \href
  {https://ui.adsabs.harvard.edu/abs/2026arXiv260121728C} {p. arXiv:2601.21728}

\bibitem[\protect\citeauthoryear{{Cookson}, {Banik}, {El-Badry}, {Sutherland},
  {Penoyre}, {Pittordis}  \& {Clarke}}{{Cookson} et~al.}{2026}]{Cookson2026}
{Cookson} S.~A.,  {Banik} I.,  {El-Badry} K.,  {Sutherland} W.,  {Penoyre} Z.,
  {Pittordis} C.,   {Clarke} C.~J.,  2026, \mn@doi [\mnras]
  {10.1093/mnras/stag342}, \href
  {https://ui.adsabs.harvard.edu/abs/2026MNRAS.547..stag342} {547, stag342}

\bibitem[\protect\citeauthoryear{{Creevey} et~al.,}{{Creevey}
  et~al.}{2023}]{Gaiaflame2}
{Creevey} O.~L.,  et~al., 2023, \mn@doi [\aap] {10.1051/0004-6361/202243688},
  \href {https://ui.adsabs.harvard.edu/abs/2023A&A...674A..26C} {674, A26}

\bibitem[\protect\citeauthoryear{{D'Odorico}, {Cristiani}, {Dekker}, {Hill},
  {Kaufer}, {Kim}  \& {Primas}}{{D'Odorico} et~al.}{2000}]{Dodorico2000UVES}
{D'Odorico} S.,  {Cristiani} S.,  {Dekker} H.,  {Hill} V.,  {Kaufer} A.,  {Kim}
  T.,   {Primas} F.,  2000, in {Bergeron} J.,  ed.,  Society of Photo-Optical
  Instrumentation Engineers (SPIE) Conference Series Vol. 4005, Discoveries and
  Research Prospects from 8- to 10-Meter-Class Telescopes. pp 121--130,
  \mn@doi{10.1117/12.390133}

\bibitem[\protect\citeauthoryear{{Desmond}, {Hees}  \& {Famaey}}{{Desmond}
  et~al.}{2024}]{Desmond2024}
{Desmond} H.,  {Hees} A.,   {Famaey} B.,  2024, \mn@doi [\mnras]
  {10.1093/mnras/stae955}, \href
  {https://ui.adsabs.harvard.edu/abs/2024MNRAS.530.1781D} {530, 1781}

\bibitem[\protect\citeauthoryear{{Di Cintio}, {Brook}, {Macci{\`o}}, {Stinson},
  {Knebe}, {Dutton}  \& {Wadsley}}{{Di Cintio} et~al.}{2014}]{DiCinito2014}
{Di Cintio} A.,  {Brook} C.~B.,  {Macci{\`o}} A.~V.,  {Stinson} G.~S.,  {Knebe}
  A.,  {Dutton} A.~A.,   {Wadsley} J.,  2014, \mn@doi [\mnras]
  {10.1093/mnras/stt1891}, \href
  {https://ui.adsabs.harvard.edu/abs/2014MNRAS.437..415D} {437, 415}

\bibitem[\protect\citeauthoryear{{Eisenstein} et~al.,}{{Eisenstein}
  et~al.}{2005}]{Eisenstein2005}
{Eisenstein} D.~J.,  et~al., 2005, \mn@doi [\apj] {10.1086/466512}, \href
  {https://ui.adsabs.harvard.edu/abs/2005ApJ...633..560E} {633, 560}

\bibitem[\protect\citeauthoryear{{El-Badry}}{{El-Badry}}{2019}]{El-badry2019}
{El-Badry} K.,  2019, \mn@doi [\mnras] {10.1093/mnras/sty3109}, \href
  {https://ui.adsabs.harvard.edu/abs/2019MNRAS.482.5018E} {482, 5018}

\bibitem[\protect\citeauthoryear{{El-Badry} \& {Rix}}{{El-Badry} \&
  {Rix}}{2018}]{Elbadry2018}
{El-Badry} K.,  {Rix} H.-W.,  2018, \mn@doi [\mnras] {10.1093/mnras/sty2186},
  \href {https://ui.adsabs.harvard.edu/abs/2018MNRAS.480.4884E} {480, 4884}

\bibitem[\protect\citeauthoryear{{El-Badry}, {Rix}  \& {Heintz}}{{El-Badry}
  et~al.}{2021}]{ElBadry2021}
{El-Badry} K.,  {Rix} H.-W.,   {Heintz} T.~M.,  2021, \mn@doi [\mnras]
  {10.1093/mnras/stab323}, \href
  {https://ui.adsabs.harvard.edu/abs/2021MNRAS.506.2269E} {506, 2269}

\bibitem[\protect\citeauthoryear{{Famaey} \& {Binney}}{{Famaey} \&
  {Binney}}{2005}]{Famaey2005}
{Famaey} B.,  {Binney} J.,  2005, \mn@doi [\mnras]
  {10.1111/j.1365-2966.2005.09474.x}, \href
  {https://ui.adsabs.harvard.edu/abs/2005MNRAS.363..603F} {363, 603}

\bibitem[\protect\citeauthoryear{{Famaey} \& {McGaugh}}{{Famaey} \&
  {McGaugh}}{2012}]{Famaey2012}
{Famaey} B.,  {McGaugh} S.~S.,  2012, \mn@doi [Living Reviews in Relativity]
  {10.12942/lrr-2012-10}, \href
  {https://ui.adsabs.harvard.edu/abs/2012LRR....15...10F} {15, 10}

\bibitem[\protect\citeauthoryear{{Gaia Collaboration} et~al.,}{{Gaia
  Collaboration} et~al.}{2016}]{Gaia2016}
{Gaia Collaboration} et~al., 2016, \mn@doi [\aap]
  {10.1051/0004-6361/201629272}, \href
  {https://ui.adsabs.harvard.edu/abs/2016A&A...595A...1G} {595, A1}

\bibitem[\protect\citeauthoryear{{Gaia Collaboration} et~al.,}{{Gaia
  Collaboration} et~al.}{2021}]{Gaia2021}
{Gaia Collaboration} et~al., 2021, \mn@doi [\aap]
  {10.1051/0004-6361/202039657}, \href
  {https://ui.adsabs.harvard.edu/abs/2021A&A...649A...1G} {649, A1}

\bibitem[\protect\citeauthoryear{{Gaia Collaboration} et~al.,}{{Gaia
  Collaboration} et~al.}{2023}]{Gaia2023}
{Gaia Collaboration} et~al., 2023, \mn@doi [\aap]
  {10.1051/0004-6361/202243940}, \href
  {https://ui.adsabs.harvard.edu/abs/2023A&A...674A...1G} {674, A1}

\bibitem[\protect\citeauthoryear{Gelman \& Rubin}{Gelman \&
  Rubin}{1992}]{gelman1992inference}
Gelman A.,  Rubin D.~B.,  1992, Statistical science, 7, 457

\bibitem[\protect\citeauthoryear{{Gentile}, {Famaey}  \& {de Blok}}{{Gentile}
  et~al.}{2011}]{Gentile2011}
{Gentile} G.,  {Famaey} B.,   {de Blok} W.~J.~G.,  2011, \mn@doi [\aap]
  {10.1051/0004-6361/201015283}, \href
  {https://ui.adsabs.harvard.edu/abs/2011A&A...527A..76G} {527, A76}

\bibitem[\protect\citeauthoryear{{Gray}}{{Gray}}{2009}]{Gray2009}
{Gray} D.~F.,  2009, {The Observation and Analysis of Stellar Photospheres}, 3
  edn.
Cambridge University Press, Cambridge

\bibitem[\protect\citeauthoryear{{Hartman} \& {L{\'e}pine}}{{Hartman} \&
  {L{\'e}pine}}{2020}]{Hartman2020}
{Hartman} Z.~D.,  {L{\'e}pine} S.,  2020, \mn@doi [\apjs]
  {10.3847/1538-4365/ab79a6}, \href
  {https://ui.adsabs.harvard.edu/abs/2020ApJS..247...66H} {247, 66}

\bibitem[\protect\citeauthoryear{{Hernandez}}{{Hernandez}}{2023}]{Hernandez2023}
{Hernandez} X.,  2023, \mn@doi [\mnras] {10.1093/mnras/stad2306}, \href
  {https://ui.adsabs.harvard.edu/abs/2023MNRAS.525.1401H} {525, 1401}

\bibitem[\protect\citeauthoryear{{Hernandez}, {Jim{\'e}nez}  \&
  {Allen}}{{Hernandez} et~al.}{2012}]{Hernandez2012}
{Hernandez} X.,  {Jim{\'e}nez} M.~A.,   {Allen} C.,  2012, \mn@doi [European
  Physical Journal C] {10.1140/epjc/s10052-012-1884-6}, \href
  {https://ui.adsabs.harvard.edu/abs/2012EPJC...72.1884H} {72, 1884}

\bibitem[\protect\citeauthoryear{{Hernandez}, {Cookson}  \&
  {Cort{\'e}s}}{{Hernandez} et~al.}{2022}]{Hernandez2022}
{Hernandez} X.,  {Cookson} S.,   {Cort{\'e}s} R.~A.~M.,  2022, \mn@doi [\mnras]
  {10.1093/mnras/stab3038}, \href
  {https://ui.adsabs.harvard.edu/abs/2022MNRAS.509.2304H} {509, 2304}

\bibitem[\protect\citeauthoryear{{Hernandez}, {Verteletskyi}, {Nasser}  \&
  {Aguayo-Ortiz}}{{Hernandez} et~al.}{2024a}]{Hernandez2024a}
{Hernandez} X.,  {Verteletskyi} V.,  {Nasser} L.,   {Aguayo-Ortiz} A.,  2024a,
  \mnras, 527, L11

\bibitem[\protect\citeauthoryear{{Hernandez}, {Verteletskyi}, {Nasser}  \&
  {Aguayo-Ortiz}}{{Hernandez} et~al.}{2024b}]{Hernandez2024}
{Hernandez} X.,  {Verteletskyi} V.,  {Nasser} L.,   {Aguayo-Ortiz} A.,  2024b,
  \mn@doi [\mnras] {10.1093/mnras/stad3446}, \href
  {https://ui.adsabs.harvard.edu/abs/2024MNRAS.528.4720H} {528, 4720}

\bibitem[\protect\citeauthoryear{{Hernandez}, {Chae}  \&
  {Aguayo-Ortiz}}{{Hernandez} et~al.}{2024c}]{Hernandez2024review}
{Hernandez} X.,  {Chae} K.-H.,   {Aguayo-Ortiz} A.,  2024c, \mn@doi [\mnras]
  {10.1093/mnras/stae1823}, \href
  {https://ui.adsabs.harvard.edu/abs/2024MNRAS.533..729H} {533, 729}

\bibitem[\protect\citeauthoryear{Hoffman, Gelman  et~al.}{Hoffman
  et~al.}{2014}]{hoffman2014no}
Hoffman M.~D.,  Gelman A.,   et~al., 2014, J. Mach. Learn. Res., 15, 1593

\bibitem[\protect\citeauthoryear{{Hopkins}}{{Hopkins}}{2018}]{Hopkins2018}
{Hopkins} A.~M.,  2018, \mn@doi [\pasa] {10.1017/pasa.2018.29}, \href
  {https://ui.adsabs.harvard.edu/abs/2018PASA...35...39H} {35, e039}

\bibitem[\protect\citeauthoryear{{Hwang}, {Ting}  \& {Zakamska}}{{Hwang}
  et~al.}{2022a}]{Hwang2022eccentricity}
{Hwang} H.-C.,  {Ting} Y.-S.,   {Zakamska} N.~L.,  2022a, \mn@doi [\mnras]
  {10.1093/mnras/stac675}, \href
  {https://ui.adsabs.harvard.edu/abs/2022MNRAS.512.3383H} {512, 3383}

\bibitem[\protect\citeauthoryear{{Hwang}, {El-Badry}, {Rix}, {Hamilton}, {Ting}
   \& {Zakamska}}{{Hwang} et~al.}{2022b}]{Hwang2022extreme-eccentric}
{Hwang} H.-C.,  {El-Badry} K.,  {Rix} H.-W.,  {Hamilton} C.,  {Ting} Y.-S.,
  {Zakamska} N.~L.,  2022b, \mn@doi [\apjl] {10.3847/2041-8213/ac7c70}, \href
  {https://ui.adsabs.harvard.edu/abs/2022ApJ...933L..32H} {933, L32}

\bibitem[\protect\citeauthoryear{{Katz} et~al.,}{{Katz}
  et~al.}{2023}]{Katz2023}
{Katz} D.,  et~al., 2023, \mn@doi [\aap] {10.1051/0004-6361/202244220}, \href
  {https://ui.adsabs.harvard.edu/abs/2023A&A...674A...5K} {674, A5}

\bibitem[\protect\citeauthoryear{{Kumar}, {Carroll}, {Hartikainen}  \&
  {Martin}}{{Kumar} et~al.}{2019}]{arviz}
{Kumar} R.,  {Carroll} C.,  {Hartikainen} A.,   {Martin} O.,  2019, \mn@doi
  [The Journal of Open Source Software] {10.21105/joss.01143}, \href
  {https://ui.adsabs.harvard.edu/abs/2019JOSS....4.1143K} {4, 1143}

\bibitem[\protect\citeauthoryear{{Langford} et~al.,}{{Langford}
  et~al.}{2025}]{Langrford2024rv}
{Langford} Z.,  et~al., 2025, \mn@doi [\pasp] {10.1088/1538-3873/ae1640}, 137,
  114503

\bibitem[\protect\citeauthoryear{{Lelli}, {McGaugh}, {Schombert}  \&
  {Pawlowski}}{{Lelli} et~al.}{2016}]{Lelli2016}
{Lelli} F.,  {McGaugh} S.~S.,  {Schombert} J.~M.,   {Pawlowski} M.~S.,  2016,
  \mn@doi [\apjl] {10.3847/2041-8205/827/1/L19}, \href
  {https://ui.adsabs.harvard.edu/abs/2016ApJ...827L..19L} {827, L19}

\bibitem[\protect\citeauthoryear{{Liu} et~al.,}{{Liu} et~al.}{2024}]{C3POII}
{Liu} F.,  et~al., 2024, \mn@doi [\nat] {10.1038/s41586-024-07091-y}, \href
  {https://ui.adsabs.harvard.edu/abs/2024Natur.627..501L} {627, 501}

\bibitem[\protect\citeauthoryear{{McGaugh}}{{McGaugh}}{2011}]{McGaugh2011MOND}
{McGaugh} S.~S.,  2011, \mn@doi [\prl] {10.1103/PhysRevLett.106.121303}, \href
  {https://ui.adsabs.harvard.edu/abs/2011PhRvL.106l1303M} {106, 121303}

\bibitem[\protect\citeauthoryear{{McGaugh}}{{McGaugh}}{2016}]{Mcgaugh2016}
{McGaugh} S.~S.,  2016, \mn@doi [\apjl] {10.3847/2041-8205/832/1/L8}, \href
  {https://ui.adsabs.harvard.edu/abs/2016ApJ...832L...8M} {832, L8}

\bibitem[\protect\citeauthoryear{{McGaugh}, {Schombert}, {Bothun}  \& {de
  Blok}}{{McGaugh} et~al.}{2000}]{McGaugh2000}
{McGaugh} S.~S.,  {Schombert} J.~M.,  {Bothun} G.~D.,   {de Blok} W. J.~G.,
  2000, \apjl, 533, L99

\bibitem[\protect\citeauthoryear{{Mel{\'e}ndez}, {Asplund}, {Gustafsson}  \&
  {Yong}}{{Mel{\'e}ndez} et~al.}{2009}]{Melendez2009}
{Mel{\'e}ndez} J.,  {Asplund} M.,  {Gustafsson} B.,   {Yong} D.,  2009, \mn@doi
  [\apjl] {10.1088/0004-637X/704/1/L66}, \href
  {https://ui.adsabs.harvard.edu/abs/2009ApJ...704L..66M} {704, L66}

\bibitem[\protect\citeauthoryear{{Meunier}, {Lagrange}, {Borgniet}  \&
  {Rieutord}}{{Meunier} et~al.}{2017}]{Meunier2017}
{Meunier} N.,  {Lagrange} A.-M.,  {Borgniet} S.,   {Rieutord} M.,  2017,
  \mn@doi [\aap] {10.1051/0004-6361/201630328}, 607, A6

\bibitem[\protect\citeauthoryear{{Milgrom}}{{Milgrom}}{1983a}]{Milgrom1983}
{Milgrom} M.,  1983a, \mn@doi [\apj] {10.1086/161131}, \href
  {https://ui.adsabs.harvard.edu/abs/1983ApJ...270..371M} {270, 371}

\bibitem[\protect\citeauthoryear{{Milgrom}}{{Milgrom}}{1983b}]{Milgrom1983b}
{Milgrom} M.,  1983b, \mn@doi [\apj] {10.1086/161132}, \href
  {https://ui.adsabs.harvard.edu/abs/1983ApJ...270..384M} {270, 384}

\bibitem[\protect\citeauthoryear{{Nelson}, {Ting}, {Hawkins}, {Ji}, {Kamdar}
  \& {El-Badry}}{{Nelson} et~al.}{2021}]{Nelson2021}
{Nelson} T.,  {Ting} Y.-S.,  {Hawkins} K.,  {Ji} A.,  {Kamdar} H.,   {El-Badry}
  K.,  2021, \mn@doi [\apj] {10.3847/1538-4357/ac14be}, \href
  {https://ui.adsabs.harvard.edu/abs/2021ApJ...921..118N} {921, 118}

\bibitem[\protect\citeauthoryear{{Oh}, {Price-Whelan}, {Hogg}, {Morton}  \&
  {Spergel}}{{Oh} et~al.}{2017}]{Oh2017}
{Oh} S.,  {Price-Whelan} A.~M.,  {Hogg} D.~W.,  {Morton} T.~D.,   {Spergel}
  D.~N.,  2017, \mn@doi [\aj] {10.3847/1538-3881/aa6ffd}, \href
  {https://ui.adsabs.harvard.edu/abs/2017AJ....153..257O} {153, 257}

\bibitem[\protect\citeauthoryear{{Oort} \& {Rougoor}}{{Oort} \&
  {Rougoor}}{1960}]{Oort1960}
{Oort} J.~H.,  {Rougoor} G.~W.,  1960, \mn@doi [\mnras]
  {10.1093/mnras/121.2.171}, \href
  {https://ui.adsabs.harvard.edu/abs/1960MNRAS.121..171O} {121, 171}

\bibitem[\protect\citeauthoryear{{Pflamm-Altenburg}}{{Pflamm-Altenburg}}{2025}]{PflammAltenburg2025}
{Pflamm-Altenburg} J.,  2025, \mn@doi [\aap] {10.1051/0004-6361/202553619},
  \href {https://ui.adsabs.harvard.edu/abs/2025A&A...703A..68P} {703, A68}

\bibitem[\protect\citeauthoryear{{Pittordis} \& {Sutherland}}{{Pittordis} \&
  {Sutherland}}{2018}]{Pittordis2018}
{Pittordis} C.,  {Sutherland} W.,  2018, \mn@doi [\mnras]
  {10.1093/mnras/sty1578}, \href
  {https://ui.adsabs.harvard.edu/abs/2018MNRAS.480.1778P} {480, 1778}

\bibitem[\protect\citeauthoryear{{Pittordis} \& {Sutherland}}{{Pittordis} \&
  {Sutherland}}{2023}]{Pittordis2023}
{Pittordis} C.,  {Sutherland} W.,  2023, \mn@doi [The Open Journal of
  Astrophysics] {10.21105/astro.2205.02846}, \href
  {https://ui.adsabs.harvard.edu/abs/2023OJAp....6E...4P} {6, 4}

\bibitem[\protect\citeauthoryear{{Planck Collaboration} et~al.,}{{Planck
  Collaboration} et~al.}{2020a}]{Planck2020a}
{Planck Collaboration} et~al., 2020a, \mn@doi [\aap]
  {10.1051/0004-6361/201833880}, \href
  {https://ui.adsabs.harvard.edu/abs/2020A&A...641A...1P} {641, A1}

\bibitem[\protect\citeauthoryear{{Planck Collaboration} et~al.,}{{Planck
  Collaboration} et~al.}{2020b}]{Planck2020b}
{Planck Collaboration} et~al., 2020b, \mn@doi [\aap]
  {10.1051/0004-6361/201833910}, \href
  {https://ui.adsabs.harvard.edu/abs/2020A&A...641A...6P} {641, A6}

\bibitem[\protect\citeauthoryear{{Pontzen} \& {Governato}}{{Pontzen} \&
  {Governato}}{2012}]{Pontzen2012}
{Pontzen} A.,  {Governato} F.,  2012, \mn@doi [\mnras]
  {10.1111/j.1365-2966.2012.20571.x}, \href
  {https://ui.adsabs.harvard.edu/abs/2012MNRAS.421.3464P} {421, 3464}

\bibitem[\protect\citeauthoryear{Rao et~al.}{Rao
  et~al.}{1945}]{rao1945information}
Rao C.~R.,  et~al., 1945, Bull. Calcutta Math. Soc, 37, 81

\bibitem[\protect\citeauthoryear{{Rubin} \& {Ford}}{{Rubin} \&
  {Ford}}{1970}]{Rubin1970}
{Rubin} V.~C.,  {Ford} Jr. W.~K.,  1970, \mn@doi [\apj] {10.1086/150317}, \href
  {https://ui.adsabs.harvard.edu/abs/1970ApJ...159..379R} {159, 379}

\bibitem[\protect\citeauthoryear{{Rubin}, {Ford}  \& {Thonnard}}{{Rubin}
  et~al.}{1980}]{Rubin1980}
{Rubin} V.~C.,  {Ford} Jr. W.~K.,   {Thonnard} N.,  1980, \mn@doi [\apj]
  {10.1086/158003}, \href
  {https://ui.adsabs.harvard.edu/abs/1980ApJ...238..471R} {238, 471}

\bibitem[\protect\citeauthoryear{Saad \& Ting}{Saad \&
  Ting}{2026}]{SaadTing2026}
Saad S.~M.,  Ting Y.-S.,  2026, \mn@doi [The Open Journal of Astrophysics]
  {10.33232/001c.170182}, 9

\bibitem[\protect\citeauthoryear{{Sacco} et~al.,}{{Sacco}
  et~al.}{2014}]{Gaiaflame}
{Sacco} G.~G.,  et~al., 2014, \mn@doi [\aap] {10.1051/0004-6361/201423619},
  \href {https://ui.adsabs.harvard.edu/abs/2014A&A...565A.113S} {565, A113}

\bibitem[\protect\citeauthoryear{{Saglia}, {Pasquini}, {Patat}, {Ludwig},
  {Giribaldi}, {Leao}, {de Medeiros}  \& {Murphy}}{{Saglia}
  et~al.}{2025}]{Saglia2025}
{Saglia} R.,  {Pasquini} L.,  {Patat} F.,  {Ludwig} H.-G.,  {Giribaldi} R.,
  {Leao} I.,  {de Medeiros} J.~R.,   {Murphy} M.~T.,  2025, \mn@doi [\aap]
  {10.1051/0004-6361/202555115}, \href
  {https://ui.adsabs.harvard.edu/abs/2025A&A...699A.151S} {699, A151}

\bibitem[\protect\citeauthoryear{{Salvatier}, {Wiecki}  \&
  {Fonnesbeck}}{{Salvatier} et~al.}{2015}]{pymc}
{Salvatier} J.,  {Wiecki} T.,   {Fonnesbeck} C.,  2015, \mn@doi [arXiv
  e-prints] {10.48550/arXiv.1507.08050}, \href
  {https://ui.adsabs.harvard.edu/abs/2015arXiv150708050S} {p. arXiv:1507.08050}

\bibitem[\protect\citeauthoryear{{Sanders}}{{Sanders}}{2003}]{Sanders2003}
{Sanders} R.~H.,  2003, \mn@doi [\mnras] {10.1046/j.1365-8711.2003.06596.x},
  \href {https://ui.adsabs.harvard.edu/abs/2003MNRAS.342..901S} {342, 901}

\bibitem[\protect\citeauthoryear{{Sanders}}{{Sanders}}{2015}]{Sanders2015MONDext}
{Sanders} R.~H.,  2015, \mn@doi [Canadian Journal of Physics]
  {10.1139/cjp-2014-0206}, \href
  {https://ui.adsabs.harvard.edu/abs/2015CaJPh..93..126S} {93, 126}

\bibitem[\protect\citeauthoryear{{Sanders} \& {McGaugh}}{{Sanders} \&
  {McGaugh}}{2002}]{Sanders2002}
{Sanders} R.~H.,  {McGaugh} S.~S.,  2002, \mn@doi [\araa]
  {10.1146/annurev.astro.40.060401.093923}, \href
  {https://ui.adsabs.harvard.edu/abs/2002ARA&A..40..263S} {40, 263}

\bibitem[\protect\citeauthoryear{{Sofue} \& {Rubin}}{{Sofue} \&
  {Rubin}}{2001}]{Sofue2001}
{Sofue} Y.,  {Rubin} V.,  2001, \mn@doi [\araa]
  {10.1146/annurev.astro.39.1.137}, \href
  {https://ui.adsabs.harvard.edu/abs/2001ARA&A..39..137S} {39, 137}

\bibitem[\protect\citeauthoryear{{Springel} et~al.,}{{Springel}
  et~al.}{2005}]{Springel2005}
{Springel} V.,  et~al., 2005, \mn@doi [\nat] {10.1038/nature03597}, \href
  {https://ui.adsabs.harvard.edu/abs/2005Natur.435..629S} {435, 629}

\bibitem[\protect\citeauthoryear{{Sun} et~al.,}{{Sun} et~al.}{2025a}]{C3POIII}
{Sun} Q.,  et~al., 2025a, \mn@doi [\apj] {10.3847/1538-4357/ad8dc3}, \href
  {https://ui.adsabs.harvard.edu/abs/2025ApJ...978..107S} {978, 107}

\bibitem[\protect\citeauthoryear{{Sun}, {Ting}, {Anthony-Twarog}, {Twarog},
  {Liu}  \& {Lu}}{{Sun} et~al.}{2025b}]{C3POV}
{Sun} Q.,  {Ting} Y.-S.,  {Anthony-Twarog} B.~J.,  {Twarog} B.~A.,  {Liu} F.,
  {Lu} Y.,  2025b, \mn@doi [\apj] {10.3847/1538-4357/adfccd}, \href
  {https://ui.adsabs.harvard.edu/abs/2025ApJ...991..185S} {991, 185}

\bibitem[\protect\citeauthoryear{{Tian}, {El-Badry}, {Rix}  \& {Gould}}{{Tian}
  et~al.}{2020}]{Tian2020}
{Tian} H.-J.,  {El-Badry} K.,  {Rix} H.-W.,   {Gould} A.,  2020, \mn@doi
  [\apjs] {10.3847/1538-4365/ab54c4}, \href
  {https://ui.adsabs.harvard.edu/abs/2020ApJS..246....4T} {246, 4}

\bibitem[\protect\citeauthoryear{{Vogt} et~al.,}{{Vogt}
  et~al.}{1994}]{Vogt1994HiRes}
{Vogt} S.~S.,  et~al., 1994, in {Crawford} D.~L.,  {Craine} E.~R.,  eds,
  Society of Photo-Optical Instrumentation Engineers (SPIE) Conference Series
  Vol. 2198, Instrumentation in Astronomy VIII. p.~362,
  \mn@doi{10.1117/12.176725}

\bibitem[\protect\citeauthoryear{{Weinberg}, {Bullock}, {Governato}, {Kuzio de
  Naray}  \& {Peter}}{{Weinberg} et~al.}{2015}]{Weinberg2015}
{Weinberg} D.~H.,  {Bullock} J.~S.,  {Governato} F.,  {Kuzio de Naray} R.,
  {Peter} A. H.~G.,  2015, \mn@doi [Proceedings of the National Academy of
  Science] {10.1073/pnas.1308716112}, \href
  {https://ui.adsabs.harvard.edu/abs/2015PNAS..11212249W} {112, 12249}

\bibitem[\protect\citeauthoryear{{Wetzel}, {Hopkins}, {Kim},
  {Faucher-Gigu{\`e}re}, {Kere{\v{s}}}  \& {Quataert}}{{Wetzel}
  et~al.}{2016}]{Wetzel2016}
{Wetzel} A.~R.,  {Hopkins} P.~F.,  {Kim} J.-h.,  {Faucher-Gigu{\`e}re} C.-A.,
  {Kere{\v{s}}} D.,   {Quataert} E.,  2016, \mn@doi [\apjl]
  {10.3847/2041-8205/827/2/L23}, \href
  {https://ui.adsabs.harvard.edu/abs/2016ApJ...827L..23W} {827, L23}

\bibitem[\protect\citeauthoryear{{Yong} et~al.,}{{Yong} et~al.}{2023}]{C3POI}
{Yong} D.,  et~al., 2023, \mn@doi [\mnras] {10.1093/mnras/stad2679}, \href
  {https://ui.adsabs.harvard.edu/abs/2023MNRAS.526.2181Y} {526, 2181}

\bibitem[\protect\citeauthoryear{{Yu} et~al.,}{{Yu} et~al.}{2025}]{C3POIV}
{Yu} J.,  et~al., 2025, \mn@doi [\mnras] {10.1093/mnras/staf436}, \href
  {https://ui.adsabs.harvard.edu/abs/2025MNRAS.538.2408Y} {538, 2408}

\bibitem[\protect\citeauthoryear{{Zhao} \& {Famaey}}{{Zhao} \&
  {Famaey}}{2006}]{Zhao2006}
{Zhao} H.~S.,  {Famaey} B.,  2006, \mn@doi [\apjl] {10.1086/500805}, \href
  {https://ui.adsabs.harvard.edu/abs/2006ApJ...638L...9Z} {638, L9}

\bibitem[\protect\citeauthoryear{{Zwicky}}{{Zwicky}}{1937}]{Zwicky1937}
{Zwicky} F.,  1937, \mn@doi [\apj] {10.1086/143864}, \href
  {https://ui.adsabs.harvard.edu/abs/1937ApJ....86..217Z} {86, 217}

\makeatother
\end{thebibliography}

\appendix
\section{Model Posteriors} \label{appendix:corner}
\begin{figure*}[ht]
    \centering
    \begin{overpic}[width=\textwidth]{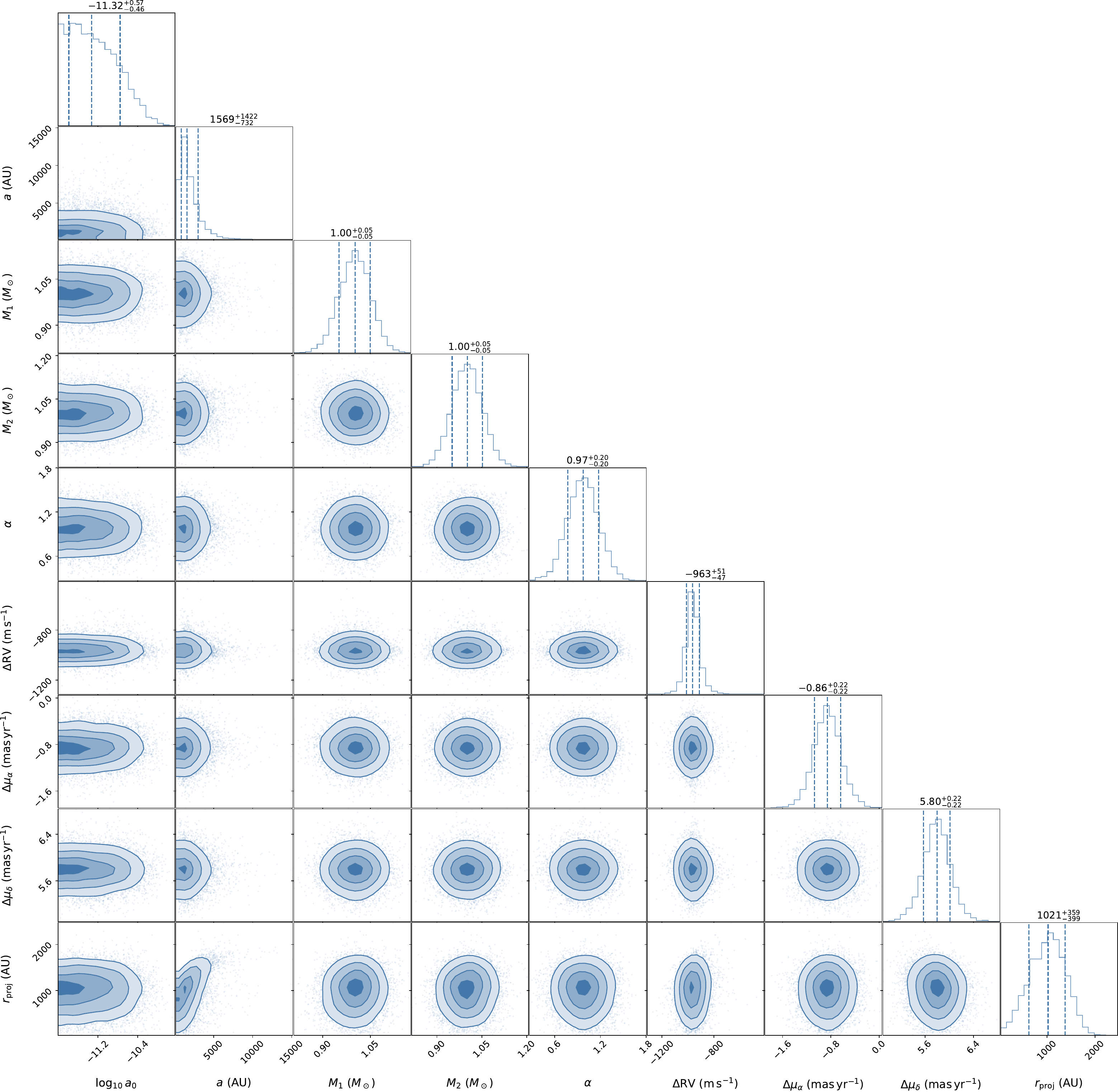}
    \end{overpic}
    \caption{Same as Figure~\ref{fig:corner94}, for binary pair Gaia 6410908065532874368 and Gaia 6410908065532874240 with $b=1$.}
    \label{fig:corner31}
\end{figure*}

\begin{figure*}[ht]
    \centering
    \begin{overpic}[width=\textwidth]{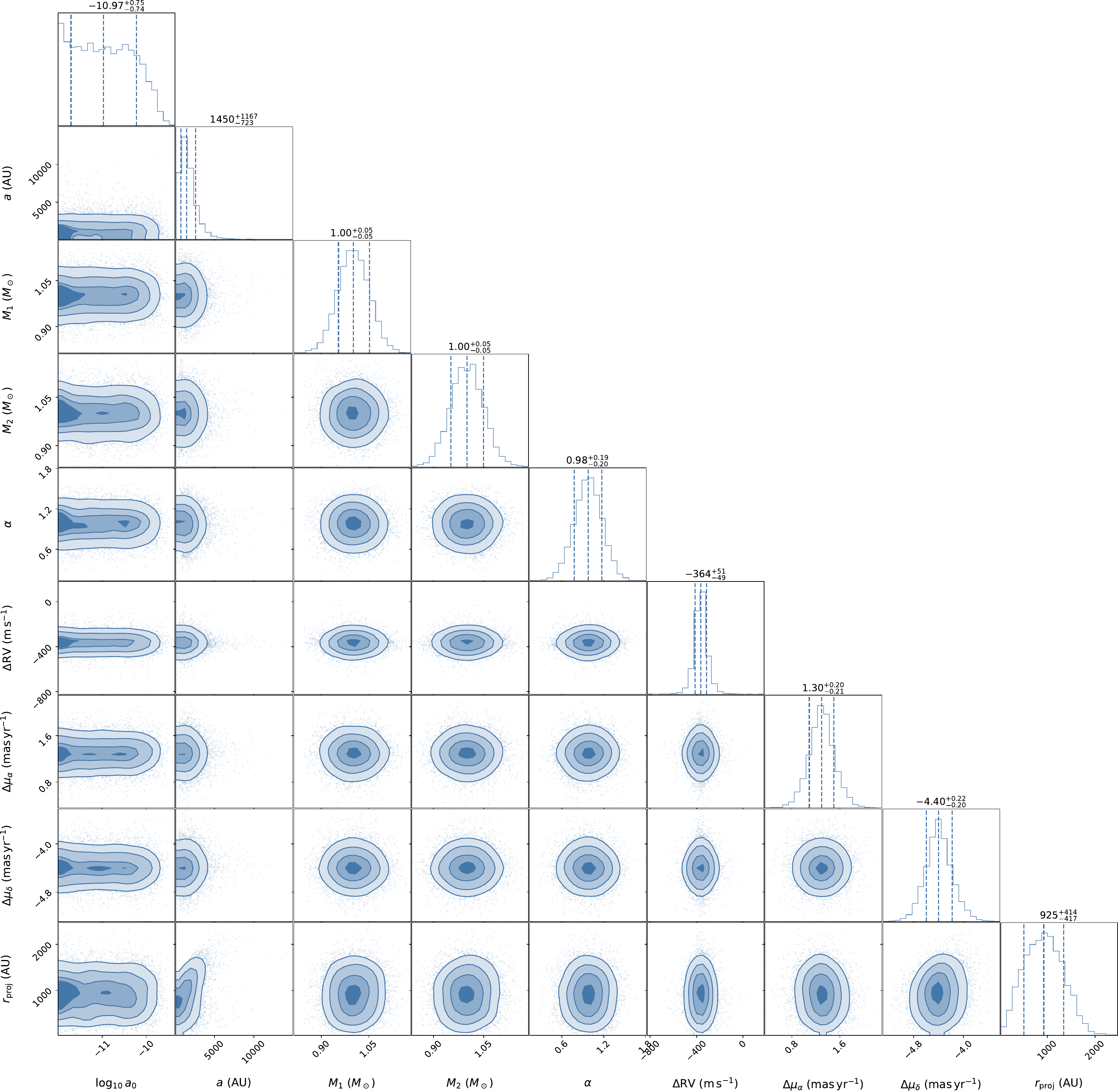}
    \end{overpic}
    \caption{Same as Figure~\ref{fig:corner94}, for the same binary pair but for $b=2$.}
    \label{fig:corner94b2}
\end{figure*}

\begin{figure*}[ht]
    \centering
    \begin{overpic}[width=\textwidth]{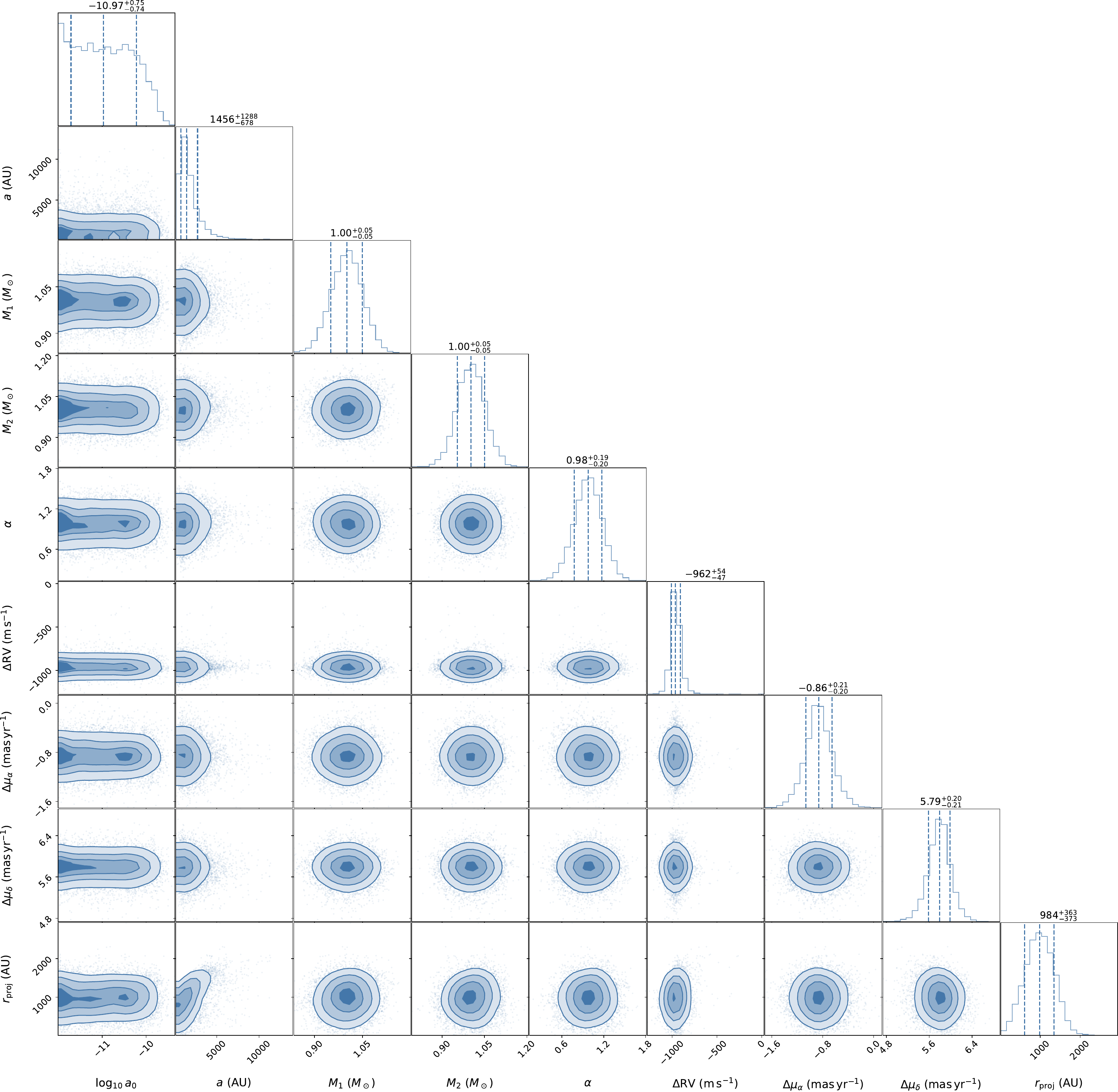}
    \end{overpic}
    \caption{Same as Figure~\ref{fig:corner31}, for the same binary pair but for $b=2$.}
    \label{fig:corner31b2}
\end{figure*}

To demonstrate the consistency of our inference across different systems and interpolating functions, we include additional corner plots.
Figure~\ref{fig:corner31} shows the posterior for a second representative binary pair with b = 1. Figures~\ref{fig:corner94b2} and~\ref{fig:corner31b2} show the same two systems as Figures~\ref{fig:corner94} and~\ref{fig:corner31}, but analyzed with b = 2.

\section{Prior Sensitivity Analysis} \label{appendix:prior_sensitivity}

\begin{figure*}
    \centering
    \begin{minipage}[b]{0.48\textwidth}
        \centering
        \includegraphics[width=\textwidth]{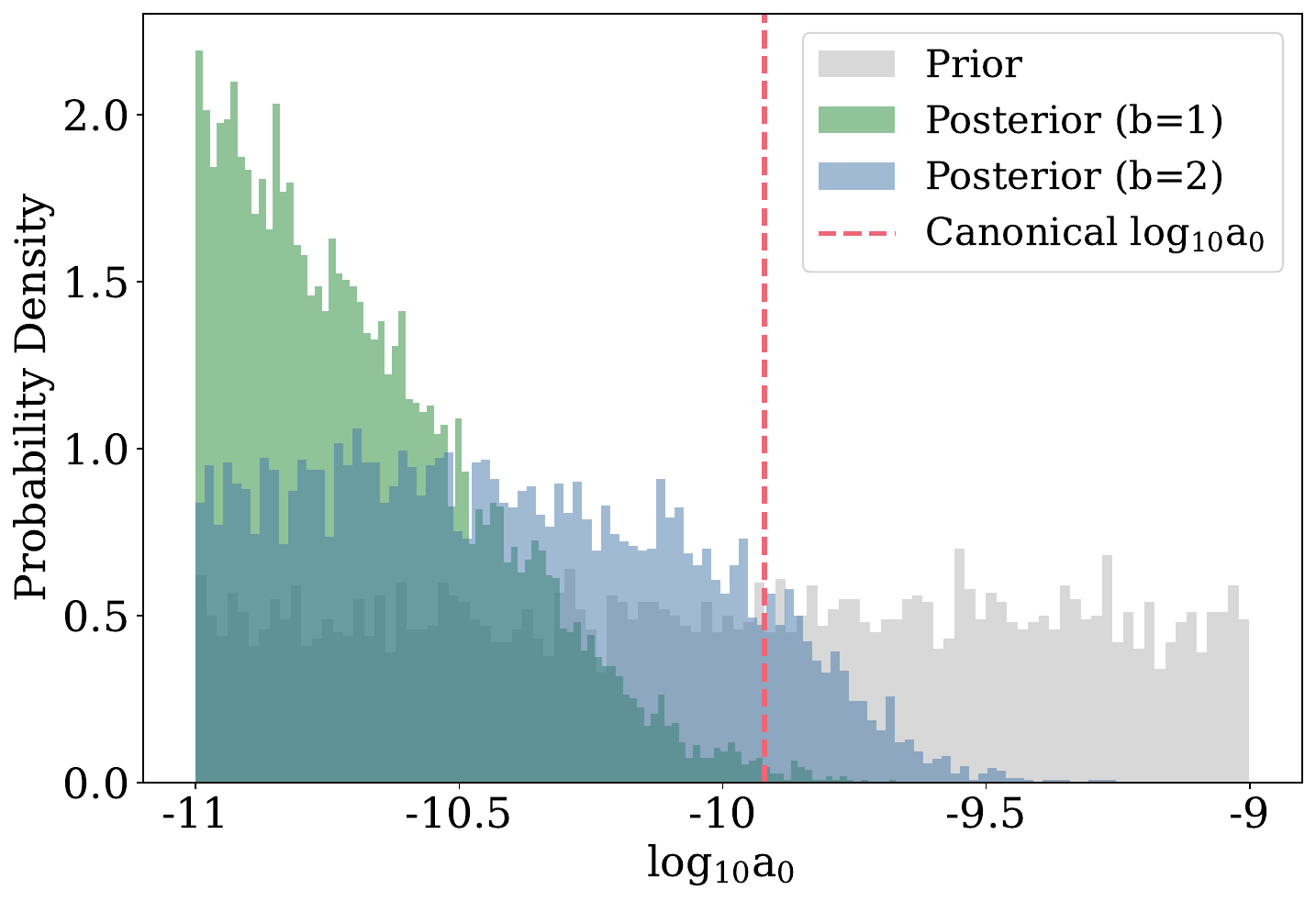}
    \end{minipage}
    \hfill
    \begin{minipage}[b]{0.48\textwidth}
        \centering
        \includegraphics[width=\textwidth]{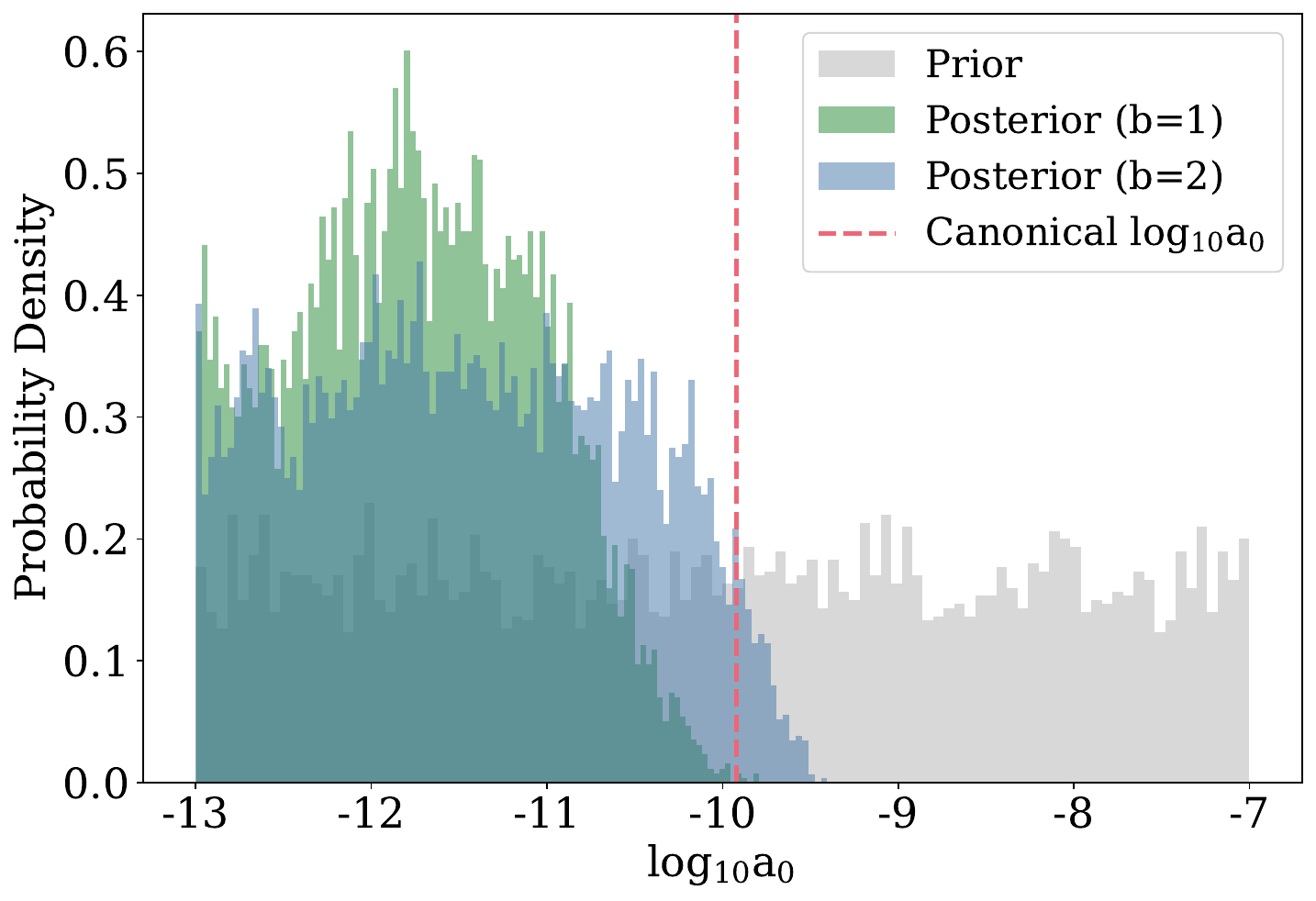}
    \end{minipage}
    \caption{Posterior distributions for $\log_{10} a_0$ using alternative priors. \textit{Left:} Narrower prior $\mathcal{U}(-11, -9)$. \textit{Right:} Wider prior $\mathcal{U}(-13, -7)$. In each panel, the gray histogram shows the uniform prior and the vertical dashed red line marks the canonical MOND value $\log_{10} a_0 = -9.92$. This is similar to Figure~\ref{fig:posterior_b12} for the prior $\mathcal{U}(-12, -8)$.}
    \label{fig:posterior_prior_sensitivity}
\end{figure*}

To further verify the dependence of our results to the choice of prior on $\log_{10}a_0$, we repeated our analysis with two alternative prior ranges: a narrower prior $\mathcal{U}(-11, -9)$ and a wider prior $\mathcal{U}(-13, -7)$. Table~\ref{tab:prior_sensitivity} summarizes the results for all three prior choices, and Figure~\ref{fig:posterior_prior_sensitivity} shows the corresponding posterior distributions.

\begin{table*}
\caption{Prior sensitivity analysis for $\log_{10} a_0$.}
\label{tab:prior_sensitivity}
\centering
\begin{tabular}{llcccc}
\hline
 & & & & Posterior & Equivalent  \\
 & & & & CDF & Gaussian  \\
$b$ & Prior & Median & 68\% CI & at Canonical & Exclusion \\
\hline
\multirow{3}{*}{1}
 & $\mathcal{U}(-11, -9)$ & $-10.72$ & $[-10.92, -10.41]$ & 99.5\% & $2.8\sigma$ \\
 & $\mathcal{U}(-12, -8)$ & $-11.24$ & $[-11.75, -10.65]$ & 99.4\% & $2.5\sigma$ \\
 & $\mathcal{U}(-13, -7)$ & $-11.75$ & $[-12.53, -11.00]$ & 99.9\% & $3.4\sigma$ \\
\hline
\multirow{3}{*}{2}
 & $\mathcal{U}(-11, -9)$ & $-10.45$ & $[-10.82, -10.02]$ & 89.6\% & $1.6\sigma$ \\
 & $\mathcal{U}(-12, -8)$ & $-10.83$ & $[-11.58, -10.15]$ & 92.9\% & $1.5\sigma$ \\
 & $\mathcal{U}(-13, -7)$ & $-12.49$ & $[-12.49, -10.42]$ & 96.3\% & $2.1\sigma$ \\
\hline
\end{tabular}
\end{table*}

\section{Gravity Boost Factor $\gamma$}
\label{appendix:gamma_test}

In addition to inferring $a_0$ directly, we performed a supplementary
test in which we inferred an effective gravity boost factor $\gamma$ for the  {57} C3PO systems. In this parameterization, the gravitational interaction is written as
$G_{\rm eff} = \gamma\,G_{\rm N}$, so that Newtonian gravity corresponds to $\gamma = 1$.
Several previous wide-binary studies have reported results in terms of $\gamma$ rather
than $a_0$ \citep{Hernandez2024a, Chae2023, Chae2024b, Chae2026, Banik2024}, and this
test allows direct comparison with those works.

\begin{figure*}
    \centering
        \includegraphics[width=0.48\textwidth]{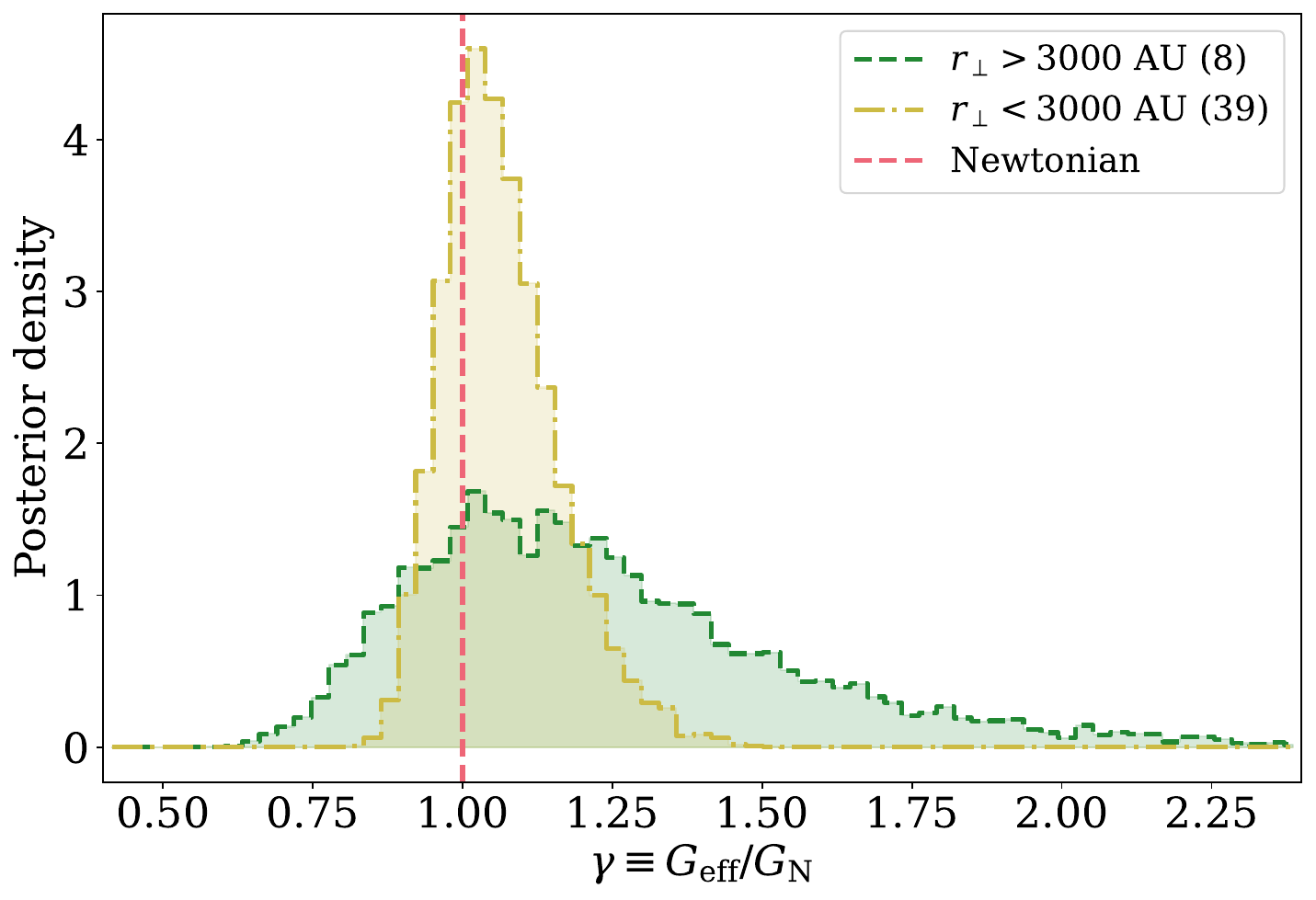}
        \hfill
        \includegraphics[width=0.48\textwidth]{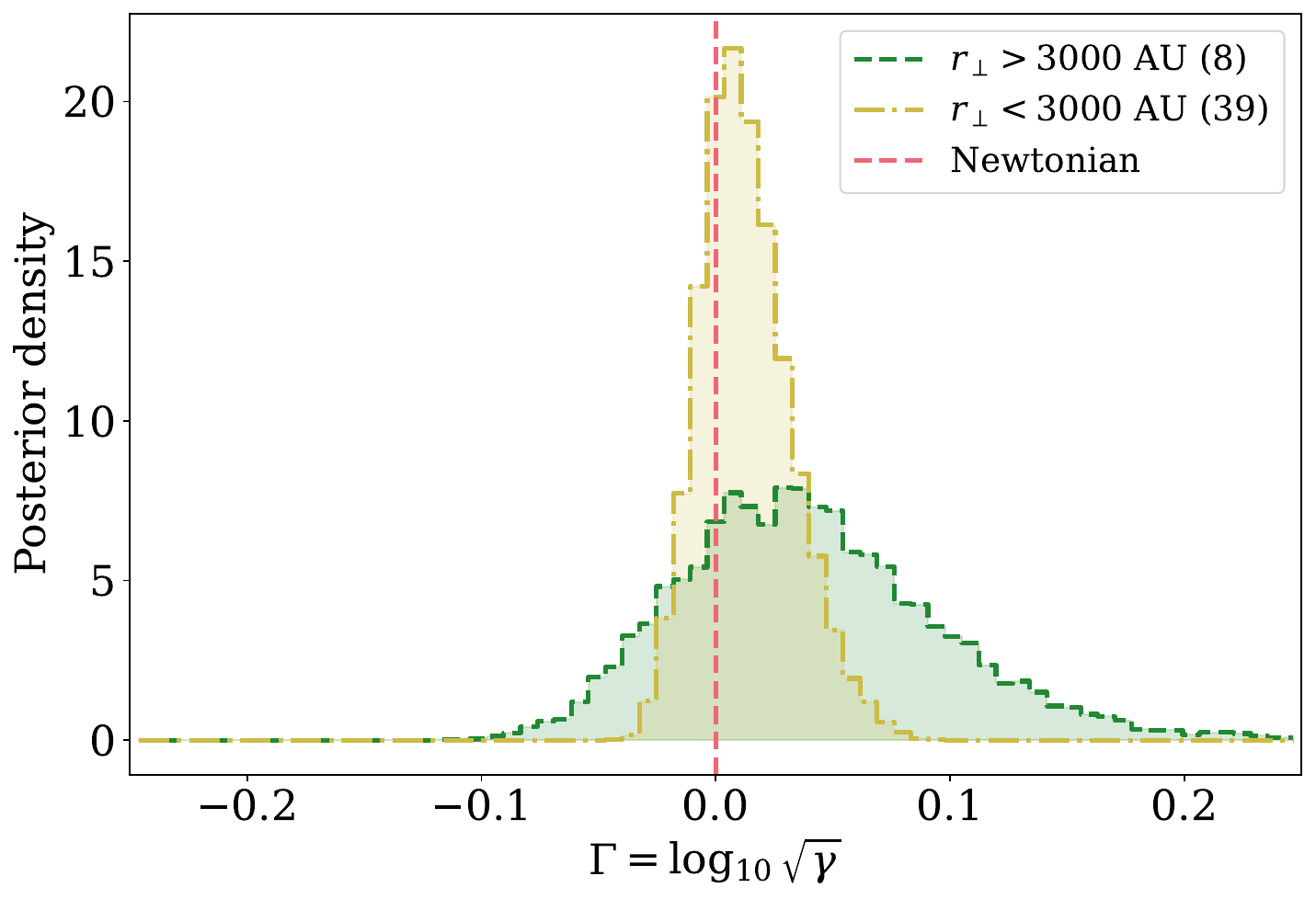}
        \caption{Posterior distributions for the gravity boost factor
        $\gamma$ (left) and the transformed parameter $\Gamma = \log_{10}\!\sqrt{\gamma}$
        (right)  {for} the  {57} systems {bound in three dimensions, split at a projected separation of 3000~AU: the 18 systems above (green, dashed) and the 39 below (orange, dash-dotted)}. The vertical dashed lines indicate the Newtonian expectation $\gamma = 1$ and $\Gamma = 0$.}
        \label{fig:gamma_posterior}
\end{figure*}

The $\gamma$-test model is identical to the main orbital framework described in
Section~\ref{sec:formulation}, except that the MOND-modified acceleration is replaced by
a simple rescaling of Newtonian gravity. The orbital velocity magnitude becomes
\begin{equation}
v = \sqrt{\frac{\gamma\,G_{\rm N}\,M_{\rm tot}}{r}} \cdot F
\label{eq:v_gamma}
\end{equation}
where $F = (1 + e^2 + 2e\cos\nu)/(1 + e\cos\nu)$ accounts for the orbital-phase
dependence in an eccentric orbit. We define $\Gamma = \log_{10}\!\sqrt{\gamma}$ and place
a uniform prior $\Gamma \sim \mathcal{U}(-1, 1)$. All other priors, likelihoods, and
inference settings are unchanged. The graphical model is shown in
Figure~\ref{fig:newton_diagram}.  {Table~\ref{tab:gamma} gives the result for the two subsamples split at a projected separation of 3000~AU, fitted separately. For the 18 systems above 3000~AU we find $\gamma = 1.17^{+0.37}_{-0.23}$ and for the 39 below $1.05^{+0.11}_{-0.08}$, both within $1\sigma$ of} Newtonian gravity ( {$\hat{R} \le 1.04$).} Figure~\ref{fig:gamma_posterior} {shows the two subsample posteriors}.

{\begin{table*}
\centering
\caption{Gravity boost factor $\gamma$ for the 57 systems bound in three dimensions, fitted separately above and below a projected separation of 3000~AU.}
\label{tab:gamma}
\begin{tabular}{lcccc}
\hline
Subset & $N$ & $\gamma$ & $P(\gamma<1)$ & $\hat{R}$ \\
\hline
$r_\perp > 3000$ AU & 18 & $1.17^{+0.37}_{-0.23}$ & 0.24 & 1.00 \\
$r_\perp < 3000$ AU & 39 & $1.05^{+0.11}_{-0.08}$ & 0.27 & 1.04 \\
\hline
\end{tabular}
\end{table*}} \begin{figure*}
    \centering
    {\usetikzlibrary{fit,positioning}

\begin{tikzpicture}[
    scale=1.3,
    shorten >=1pt,->,draw=black!70,
    neuron/.style={circle,minimum size=40,inner sep=0},
    param/.style={neuron,fill=cyan}, %
    inter/.style={neuron,fill=black!40}, %
    gauss/.style={neuron,fill=red}, %
    obs/.style={neuron,fill=yellow}, %
    beta/.style={neuron,fill=yellow}, %
    box/.style={rectangle,draw=black,rounded corners,thick,inner sep=10pt}
]

\node[param] (a)   at (0,4) {$a$};
\node[param] (alpha)   at (-1.5,4) {$\alpha$};
\node[inter] (nu)  at (-3,1) {$\nu$};
\node[param] (M)   at (-3,4) {$\phi$};
\node[param] (i)   at (3,4) {$\cos i$};
\node[param] (omega)   at (4.5,4) {$\omega$};
\node[param] (Omega)   at (4.5,2.5) {$\Omega$};

\node[param] (M1M2) at (1.5,4)
  {\parbox{1cm}{\centering $M_1$ \\ $M_2$}};

\node[inter] (e) at (-1.5,1) {$e$};
\node[inter] (rtrue) at (-1.5,-0.5) {$r_{\rm true}$};
\node[inter] (R) at (3.75,1) {$R$};
\node[inter] (vproj) at (3.75,-0.5) {$\parbox{1cm}{\centering $v_{1,\rm proj}$ \\ $v_{2,\rm proj}$}$};

\node[inter] (v1v2) at (0.75,1) {$\parbox{1cm}{\centering $v_1$ \\ $v_2$}$};
\node[inter] (rv) at (0.75,-2) {$\Delta \rm RV$};
\node[inter] (pmra) at (2.25,-2) {$\Delta \mu_{\alpha}$};
\node[inter] (pmdec) at (3.75,-2) {$\Delta \mu_{\delta}$};
\node[inter] (rproj)   at (-1.5,-2) {$r_{\rm proj}$};

\node[gauss] (uM)   at (-3,2.5) {$\mathcal{U}$};
\node[gauss] (ue)   at (-1.5,2.5) {$\mathcal{U}$};
\node[gauss] (ua)   at (0,2.5) {$\mathcal{U}$};
\node[gauss] (gM1M2)   at (1.5,2.5) {$\mathcal{N}$};
\node[gauss] (uR)   at (3,2.5) {$\mathcal{U}_{3}$};
\node[gauss] (gv1v2) at (2.25,-3.5) {$\mathcal{N}_{4}$};

\node[obs] (robs) at (-0.75,-5) {$r_{\rm proj}$};
\node[obs] (rvobs) at (0.75,-5) {$\Delta \rm RV$};
\node[obs] (pmraobs) at (2.25,-5) {$\Delta \mu_{\alpha}$};
\node[obs] (pmdecobs) at (3.75,-5) {$\Delta \mu_{\delta}$};

\draw[->] (i) -- (uR);
\draw[->] (omega) -- (uR);
\draw[->] (Omega) -- (uR);
\draw[->] (uR) -- (R);
\draw[->] (R) -- (rproj);
\draw[->] (M) -- (uM);
\draw[->] (uM) -- (nu);

\draw[->] (a) -- (ua);
\draw[->] (ua) -- (rtrue);
\draw[->] (e) -- (rtrue);
\draw[->] (alpha) -- (ue);
\draw[->] (ue) -- (e);
\draw[->] (nu) -- (rtrue);

\draw[->] (M1M2) -- (gM1M2);
\draw[->] (gM1M2) -- (v1v2);
\draw[->] (rtrue) -- (rproj);

\draw[->] (R) -- (v1v2);
\draw[->] (rtrue) -- (v1v2);

\draw[->] (v1v2) -- (vproj);
\draw[->] (R) -- (vproj);
\draw[->] (vproj) -- (rv);
\draw[->] (vproj) -- (pmra);
\draw[->] (vproj) -- (pmdec);

\draw[->] (rproj) -- (gv1v2);
\draw[->] (gv1v2) -- (robs);
\draw[->] (rv) -- (gv1v2);
\draw[->] (gv1v2) -- (rvobs);
\draw[->] (pmra) -- (gv1v2);
\draw[->] (gv1v2) -- (pmraobs);
\draw[->] (pmdec) -- (gv1v2);
\draw[->] (gv1v2) -- (pmdecobs);

\node[box, fit=(a)(e)(M)(nu)
             (M1M2)(rtrue)(i)(omega)(Omega)
             (v1v2)(rv)(pmra)(pmdec)
             (gv1v2)(robs)(rvobs)(pmraobs)(pmdecobs)] (bigbox) {};

\node[anchor=south west,font=\small] at ([xshift=10pt,yshift=10pt]bigbox.south west) {For each pair};

\node[anchor=east] at ([xshift=-0.5cm]bigbox.west) {
    \begin{tikzpicture}[every node/.style={anchor=west}, x=1cm, y=0.6cm]
        \node[param,label=right:{Parameter}] at (3,6) {};
        \node[inter,label=right:{Intermdeiate Variables}] at (3,3) {};
        \node[gauss,label=right:{Distribution}] at (3,0) {};
        \node[obs,label=right:{Observed Variables}] at (3,-3) {};
    \end{tikzpicture}
};

\end{tikzpicture}}
    \caption{Graphical representation of the hierarchical Bayesian model
    used for the $\gamma$ test. The structure is identical to the main orbital model shown
    in Figure~\ref{fig:mond_diagram}, except that the global parameter is the gravity boost
    factor $\gamma$ instead of the MOND acceleration scale $a_0$.}
    \label{fig:newton_diagram}
\end{figure*}

This result must be interpreted with care. Because most of the C3PO
systems reside in the MOND transition regime ($a_N \sim a_0$), the effective gravity boost
predicted by MOND is not expected to be constant across all binaries. In this regime, the
predicted boost depends on the interpolating function, the treatment of the external field
effect, and the acceleration of each individual system. MOND predicts $\gamma > 1$ but
with values that vary from system to system in transition regime; fitting a single global
$\gamma$ to systems spanning a range of accelerations within the transition regime will
tend to average over these varying boosts. Furthermore, the predicted boost in the
transition regime is smaller than in the deep MOND regime, so even if MOND is correct, the
C3PO sample would be expected to yield a value of $\gamma$ closer to unity than the
$\gamma \approx 1.4$--$1.6$ predicted for systems deep in the MOND regime. For these
reasons, a $\gamma$ measurement is most informative when applied to systems that reside
firmly in the deep MOND regime, where the predicted boost is approximately constant across
the sample.

On the other hand, if Newtonian gravity is the correct description, then $\gamma = 1$
should hold regardless of the acceleration regime. Our  {measurements are} consistent with this expectation,  {and the subsample above 3000~AU, which is where previous studies report a boost of 1.3 to 1.6, gives $1.17^{+0.37}_{-0.23}$}. We therefore do not regard $\gamma$ as the primary
parameter of interest for the C3PO sample, and instead focus on the direct inference of
$a_0$ as the main result of this paper.

 \section{Population-Level Orbital Parameter Distributions}
\label{appendix:population_orbital}

As a consistency check on our hierarchical model, we examine the
population-level posterior distributions of $\cos i$ and mean anomaly $M$ across all 57
systems, following the diagnostics requested by the referee. For orbits observed at random
times with isotropic orientations, the distribution of $\cos i$ should be uniform on
$[-1, 1]$ and the distribution of $M$ should be uniform on $[0, 2\pi]$.

\begin{figure*}
    \centering
    \includegraphics[width=\textwidth]{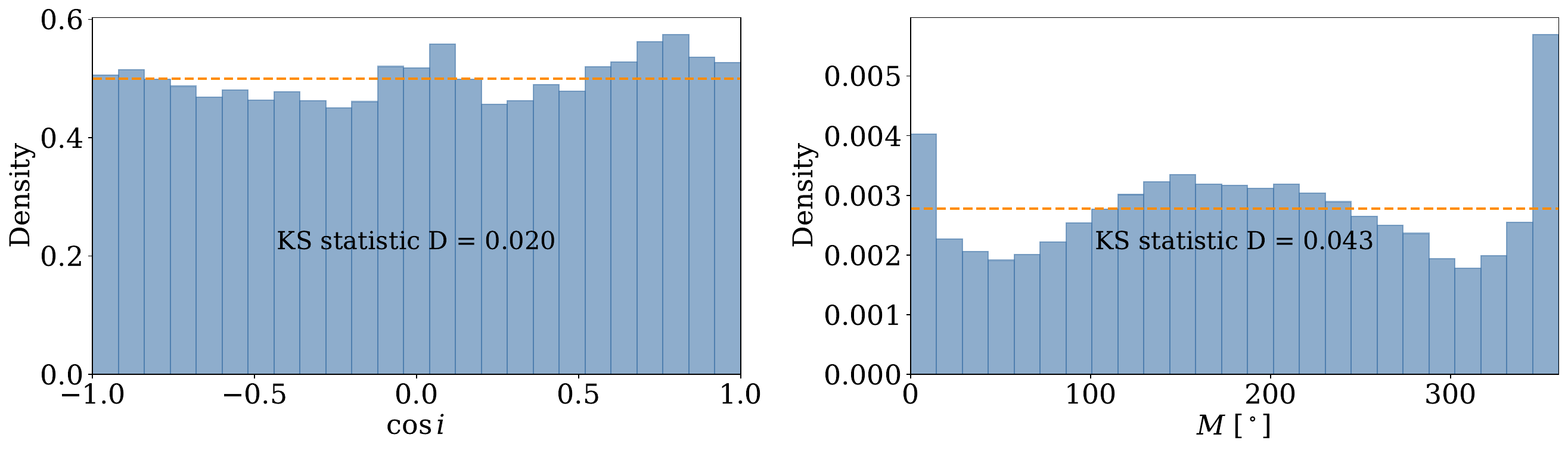}
    \caption{Population-level posterior distributions of $\cos i$ (left)
    and mean anomaly $M$ (right) across all 57 binary systems. The histograms show the
    combined posterior samples from all chains and draws. The dashed orange line indicates
    the expected uniform distribution. The Kolmogorov--Smirnov statistic $D$ is shown in
    each panel, confirming consistency with uniformity. The distribution of $\cos i$ is
    consistent with isotropy, and the distribution of $M$ is consistent with uniform phase
    sampling, as expected for gravitationally bound systems observed at random times.}
    \label{fig:population_cosi_M}
\end{figure*}

Figure~\ref{fig:population_cosi_M} shows the posterior distributions of $\cos i$ and
mean anomaly $M$ (in degrees) for all 57 systems, combining all posterior samples across
chains and draws. The dashed orange line indicates the expected uniform distribution in
each panel. We quantify the agreement using the Kolmogorov--Smirnov (KS) statistic $D$,
which measures the maximum deviation between the empirical cumulative distribution and the
theoretical uniform distribution. The KS statistic values are shown in each panel.

Both distributions are consistent with uniformity: the KS statistics are small ( {$D < 0.05$}),
indicating only minor deviations. The consistency of $\cos i$ with a uniform distribution
confirms that the model produces physically reasonable, isotropic orbital orientations.
The uniformity of $M$ confirms that the posterior sampling is consistent with gravitationally bound systems observed at random times.

{{} \section{Inferred Apoapsis Distribution and Tidal Radii} \label{appendix:apoapsis}

A necessary consistency check on any orbit fit to a wide binary is that the inferred orbit remains bound against the Galactic tide. If the inferred maximum separation exceeded the tidal radius, the solution would be unphysical. We therefore compare the posterior apoapsis of each system, $a(1+e)$, with its Jacobi radius in the solar neighbourhood,
\begin{equation}
r_{\rm J} = R_0 \left(\frac{M_{\rm tot}}{3\,M_{\rm gal}(<R_0)}\right)^{1/3},
\label{eq:jacobi}
\end{equation}
where $M_{\rm gal}(<R_0) = v_c^2 R_0/G$ with $R_0 = 8$~kpc and $v_c = 220$~km\,s$^{-1}$. For the masses in our sample this gives $r_{\rm J} \simeq 1.6$~pc, or about $3.2\times10^5$~AU.

Figure~\ref{fig:apoapsis} shows the result for the fiducial $b=1$ posterior. The median inferred apoapsis is  {1490} ~AU, with the individual systems spanning  {158 to $4.6\times10^4$} ~AU. No system in the sample has an inferred apoapsis approaching its Jacobi radius. The largest median value reaches  {14 percent} of $r_{\rm J}$, and even at the 84th percentile of the posterior the largest value is  {28 percent}. Taking the 99th percentile of every system's apoapsis posterior, none exceeds the tidal limit. The inferred orbits are therefore comfortably bound, by roughly an order of magnitude in separation, and the fits do not rely on unphysically extended configurations.

\begin{figure}
    \centering
    \includegraphics[width=0.5\textwidth]{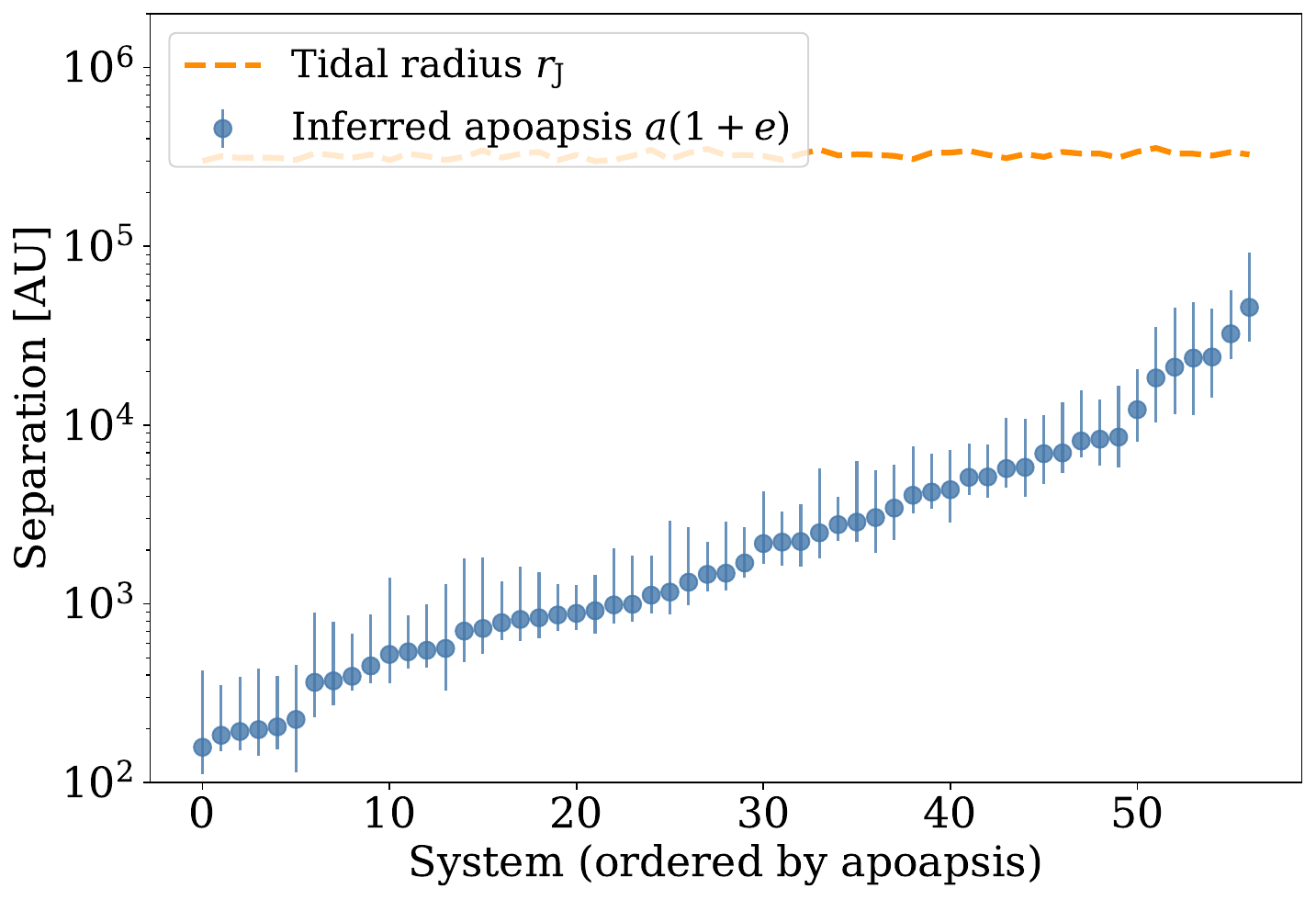}
    \caption{Inferred apoapsis $a(1+e)$ for the 57 bound systems, ordered by median value, with error bars showing the 16th--84th percentile range of the posterior. The dashed red line is the Jacobi radius of each system in the solar neighbourhood ( {Eq.} ~ {\eqref{eq:jacobi}}). Every system lies well below its tidal limit, and the most extended reaches  {14 percent} of $r_{\rm J}$ at the median and  {28 percent} at the 84th percentile.}
    \label{fig:apoapsis}
\end{figure}
{}} 

\end{document}